\documentclass[a4paper,pra,twocolumn,showpacs,superscriptaddress,groupedaddress]{revtex4}
\usepackage{ae}
\usepackage[T1]{fontenc}
\usepackage[ansinew]{inputenc}
\usepackage{amsmath}
\usepackage{amssymb}
\usepackage[caption=false]{subfig}
\usepackage{multirow}
\usepackage{array}
\usepackage[]{graphicx}
\usepackage{wrapfig}
\usepackage{makecell}
\usepackage{color}
\usepackage[colorlinks]{hyperref}
\usepackage{lscape}
\graphicspath{ {./images1/} }

\begin{document}
		\title{Nonlocality without entanglement: Party asymmetric case }
		\author{Atanu Bhunia}
		\email{atanu.bhunia31@gmail.com}
		\affiliation{Department of Applied Mathematics, University of Calcutta, 92, A.P.C. Road, Kolkata- 700009, India}
		\author{Indrani Chattopadhyay}
		\email{icappmath@caluniv.ac.in}
		\affiliation{Department of Applied Mathematics, University of Calcutta, 92, A.P.C. Road, Kolkata- 700009, India}
		\author{Debasis Sarkar}
		\email{dsarkar1x@gmail.com, dsappmath@caluniv.ac.in}
		\affiliation{Department of Applied Mathematics, University of Calcutta, 92, A.P.C. Road, Kolkata- 700009, India}
		
\begin{abstract}
	A set of orthogonal product states of a composite Hilbert space is genuinely nonlocal if the states are locally indistinguishable across any bipartition. In this work, we construct a minimal set of party asymmetry genuine nonlocal set in arbitrary large dimensional composite quantum systems $C^d\otimes C^d\otimes C^d$. We provide a local discriminating protocol by using a three qubit GHZ state as a resource. On the contrary, we observe that single-copy of two qubit Bell states provide no advantage for this discrimination task. Recently, Halder et al. [Phys. Rev. Lett. 122, 040403 (2019)], proposed the concept of strong nonlocality without entanglement and ask an open question whether there exist an incomplete strong nonlocal set or not. In [Phys. Rev. A 102, 042228 (2020)], an answer is provided by the authors. Here, we construct an incomplete party asymmetry strong nonlocal set which is more stronger than the set constructed in [Phys. Rev. A 102, 042228 (2020)] with respect to the consumption of entanglement as a resource for their respective discrimination tasks.
		\end{abstract}
		\date{\today}
		\maketitle
		\section{INTRODUCTION}
		One of the main goal of the subject quantum information theory is to figure out the structure of quantum operations which can be implemented by local operations alongwith classical communications (LOCC). Several researchers have tried to investigate the restrictions on quantum operations that can be implemented by LOCC. The local distinguishability and indistinguishability of set of quantum states is an area that attracted many researcher to understand the nature of quantum correlations present in the composite systems \cite{Bennett1999,Walgate2000,Virmani,Ghosh2001,Groisman,Walgate2002,Divincinzo,Horodecki2003,Fan2004,Ghosh2004,Nathanson2005,Watrous2005,Niset2006,Ye2007,Fan2007,Runyo2007,somsubhro2009,Feng2009,Runyo2010,Yu2012,Yang2013,Zhang2014,Yu,somsubhro2009(1),somsubhro2010,yu2014,somsubhro2014,somsubhro2016}. In case of a well known bipartite system, suppose a state is secretly chosen from a set of pre-specified orthogonal quantum states shared between two distant parties, say, Alice and Bob. Their goal is to  figure out locally the exact identity of that state and this goal is achieved by several researcher in many cases\cite{Walgate2000,Walgate2002,Fan2007,Ghosh2001,Ghosh2004,Nathanson2005}. An interesting outcome of this type of research is the local indistinguishability of pure orthogonal product states that exhibits the phenomenon of nonlocality without entanglement. The nonlocality without entanglement(NLWE) can be viewed as a new striking example of the nonequivalence between the concept of quantum entanglement and that of quantum nonlocality \cite{Zhang2015,Wang2015,Chen2015,Yang2015,Zhang2016,Xu2016(2),Zhang2016(1),Xu2016(1),Halder2019strong nonlocality,Halder2019peres set,Xzhang2017,Xu2017,Wang2017,Cohen2008,Zhang2016(3),somsubhro2018,zhang2018,Halder2018,Yuan2020,Rout2019,Rout2020,bhunia2020}. It was known that entanglement does not necessarily imply nonlocality in the sense of producing data that are incompatible with local realism. Rather, this new type of nonlocality indicates that the converse does not hold true in general. Note that, here this new type of nonlocality is not understood like the incompatibility with local realism, but instead, as the advantage of a joint measurement with respect to LOCC.
		
		Walgate et al.\cite{Walgate2000} provides that by using LOCC only any two pure orthogonal multipartite states can be perfectly distinguished. The local indistinguishability of pairwise orthogonal set of multipartite states is a signature of nonlocality showed by those states. Since entanglement is deeply attached to nonlocality, one can assume that pairwise orthogonal product states would be perfectly distinguished by LOCC. However, this assumption is wrong. Bennett et al. \cite{Bennett1999} showed that there exist a complete set of nine pure orthogonal product states in \({\mathbb {C}}^{3}\bigotimes {\mathbb {C}}^{3}\) which is
		indistinguishable by LOCC. So, the local distinguishability problem is not so simple when the number of states is more than two.   Recently, almost all the studies focused on local discrimination of bipartite quantum states, entangled or not and not much is known for multipartite systems. For tripartite quantum states, Halder et al., exhibit strong quantum nonlocality without entanglement \cite{Halder2019strong nonlocality} and constructed two explicit examples of strongly nonlocal sets of quantum states in
	   \({\mathbb {C}}^{3}\otimes {\mathbb {C}}^{3}\otimes {\mathbb {C}}^{3}\) and \({\mathbb {C}}^{4}\otimes {\mathbb {C}}^{4}\otimes {\mathbb {C}}^{4}\) quantum systems, respectively. In their opinion, strong quantum nonlocality if exists, for tripartite quantum states, they are locally irreducible in every bipartition. In another work, Rout et al.\cite{Rout2019} classified genuinely nonlocal product bases (GNPBs) of multipartite quantum systems. In \cite{Yuan2020}, the authors constructed an incomplete strong nonlocal set which was an open question in \cite{Halder2019strong nonlocality}. In \cite{Rout2019}, the authors provided entanglement assisted discrimination protocol of strong nonlocal basis and showed that it consumes greater amount of entanglement resource with respect to any other nonlocal set in $3\otimes3\otimes3$. Still it is interesting to know that whether there are strong nonlocal sets those need much entanglement resource for discrimination tasks or not.\\
	   
	   The GNPBs constructed in \cite{Rout2019} contains a large number of states about their respective dimensions and also the generalization is not proper with the higher dimensional cases. So, it is important to find genuinely nonlocal product sets (GNPS) which contains lesser number states that respects their dimension and generalize properly for higher dimensional subsystems. The nonlocal sets constructed in \cite{Yuan2020} and \cite{Halder2019strong nonlocality} are the strongest form of nonlocality till now. The only difference there, one is complete and the another is incomplete. Both of them have party symmetric structure as well as consumes same amount of entanglement for their respective discrimination tasks. So it is further an issue to find strong nonlocal set which is incomplete and needs much resource for discrimination task than previous one. In \cite{Rout2019}, authors explored  the discrimination task by using GHZ state as resource for the $GNPB_{II}(3,4)$. But their protocol do not provide advantage by using GHZ state as a resource for the discrimination task $GNPB_{II(a)}(3,3)$. They showed that two copies of two-qubit Bell states are necessary for the discrimination task of $GNPB_{II(a)}(3,3)$. In this manuscript, we have Provided that one copy of three-qubit GHZ state is sufficient to discriminate perfectly a class of state which has same characteristic as $GNPB_{II(a)}(3,3)$ in the sense of genuine nonlocality. We generalize our result for higher dimensions also.\\   
	   
	   The remaining portion of this paper is arranged as follows. In Sec.~\ref{A1} necessary definitions and other preliminary concepts
	   are presented. In Sec.~\ref{A2}, we construct a minimal set of party asymmetry genuine nonlocal set in arbitrary large dimensional quantum systems $C^d\otimes C^d\otimes C^d$. Also, in Sec.~\ref{A3} we have succeeded to construct an incomplete strong nonlocal set of orthogonal product states which is more stronger than the set constructed in \cite{Yuan2020} with respect to consumption of entanglement as a resource for their respective LOCC discrimination tasks. In Sec.~\ref{A4}, we provide a discriminating protocol by using a three qubit GHZ state as a resource. It is interesting to observe that single-copy of two qubit Bell state provides no advantages for such local discriminating tasks. Finally, the conclusion is drawn in Sec.~\ref{A5} with some open problems for further studies.
		
		\section{Preliminaries}
		\label{A1}
		In this section, we will review some definitions which are useful for our further discussions.\\
	
		$Definition~ 1.$ \cite{Halder2018} Suppose all the POVM elements of a measurement structure corresponding to a discrimination task of a given set of states are proportional to the identity matrix. Then such a measurement is not useful to extract information  for this task and is called $trivial\; measurement$. \\
		On the other hand, if not all the POVM elements of a measurement are proportional to the identity matrix then the measurement	is said to be a $nontrivial\; measurement$.\\\\
		$Definition ~2.$ \cite{Halder2018} Consider a measurement to distinguish a fixed set of pairwise orthogonal quantum states. After performing that measurement, if the post-measurement states are also	pairwise orthogonal to each other then such a measurement is said to be an $orthogonality-preserving\;\;measurement$(OPM).\\\\
		$Definition~ 3.$ \cite{Halder2019strong nonlocality} A set of orthogonal quantum states is $locally\; irreducible$ if it is not possible to eliminate one or more quantum states from the set by nontrivial orthogonality-preserving local measurements.\\\\
		$Definition~ 4.$ A set of orthogonal
		quantum states is called $locally\; indistinguishible$  if it is not possible to distinguish completely the whole set by nontrivial orthogonality-preserving local measurements.\\
		Therefore it is by definition imply that all locally irreducible states are locally indistinguishable but the converse is not true in general.\\\\
		$Definition~ 5.$ \cite{Halder2019strong nonlocality} A set of quantum states in a tripartite system is said to be $strong \;nonlocal$ if it is locally irreducible in tripartition and also locally irreducible in every bipartition.\\
		
		When sufficient entanglement is available as a resource among the parties along with their operational power of LOCC,	it is possible to perfectly discriminate a nonlocal set. Use of quantum teleportation in distinguishing orthogonal set of quantum states is an example of such protocol \cite{Ghosh2004}. In that work the authors showed that if the involved parties share sufficient amount of entanglement, then using teleportation respective parties can perfectly discriminate the states by performing suitable measurements. Now it is well understood that entanglement is a valuable resource under several operational tasks using LOCC. So, any protocol consuming less entanglement (i.e., less resource) in a specific task, is always better than other protocols achieving that same task. An example of such protocols for a class of locally indistinguishable product states was proposed by Cohen \cite{Cohen2008}. They showed that the well-known Bennett’s\cite{Bennett1999} two-qutrit nonlocal product basis (NPB) can be perfectly distinguished by using one ebit of entanglement as a resource, whereas the teleportation-based protocol requires a two-qutrit maximally entangled state, i.e., $\log_2 3$ ebit. Cohen’s result \cite{Cohen2008} motivates many researcher in identifying efficient use of entanglement in the local state discriminating problems. Motivating with the result just stated above we have tried to figure out such protocols in multipartite systems. In the next section, we have considered different problems in tripartite quantum systems where we have addressed the question of resource consumption.
		
		\section{Construction of nonlocal sets and their distinguishability issue}
		
		The twisted states \cite{Niset2006} play an important role in local distinguishability of orthogonal product states. Using entanglement as a resource to discriminate a set of locally indistinguishable states, we need sufficient ancillary subsystems \cite{Cohen2008,Rout2019}. For the discrimination task with support of a two qubit Bell state, the dimension of each ancillary subsystem required is two. To generate a discrimination protocol, sharing Bell state, respective parties chooses appropriate measurements on his/her system together with the ancillary subsystems. As well known that for a tripartite system if any two parties want to come together, then one of them has to teleport his/her system to other. Hence he/she needs a maximally entangled state to teleport. If their system dimensions greater than two, then the ancillary subspace must have dimension larger than two \cite{Cohen2008}. So the choice of measurement for Bell state assisted discrimination task must be a subset of the choice of measurement for higher dimensional maximally entangled state assisted discrimination tasks. As an example for $3\otimes3\otimes3$ system, to teleport one system to other a two qutrit maximally entangled state needed. In this case, the minimum dimension of ancillary subspace is three, say, generated by $(|0\rangle, |1\rangle, |2\rangle)$. Whereas, for a Bell state assisted discrimination task the ancillary subspace dimension is two. We summarize the above discussions by writing the following proposition.\\
		
		{\bfseries Proposition 1}. In a tripartite system if a set of orthogonal product states is distinguishable by using a two qubit Bell state shared between any two parties, then the set of states is distinguishable while the corresponding parties join together. But the converse is not true always.\\
		
		In a tripartite system, local indistinguishability in any bipartite cut say $AB|C$ is more stronger than the indistinguishability by using a two-qubit maximally entangled state shared between $A$ and $B$; i.e.,  those tripartite  sets of states which are distinguishable by using a two qubit maximally entangled state shared between any two parties, say, $A$ and $B$,  are also distinguishable in the bipartition $AB|C$ cut. Consider an example of a set of orthogonal product states (we avoid normalization constants here and throughout the paper) shared between three parties Alice, Bob and Charlie in $3 \otimes 3 \otimes 3$.\\
		
\begin{multline}
	$$
	\;\;\;\;\;\;\left|\phi_{1,2}\right\rangle=|0\pm1\rangle|0\rangle|1\rangle,\;\;\; \left|\phi_{3,4}\right\rangle=|1\rangle|0\pm 1\rangle|0\rangle\\
	\left|\phi_{5,6}\right\rangle=|0\rangle|1\rangle|0 \pm 1\rangle,\;\;\;  \left|\phi_{7,8}\right\rangle=|2\rangle|1\rangle|1 \pm 2\rangle \\
	\left|\phi_{9,10}\right\rangle=|1\rangle|1 \pm 2\rangle|2\rangle,\;\;\;  \left|\phi_{11,12}\right\rangle=|1 \pm 2\rangle|2\rangle|1\rangle \\
	\left|\phi_{13,14}\right\rangle=|0\rangle|2\rangle|1 \pm 2\rangle,\;\;\;
	 \left|\phi_{15,16}\right\rangle=|1 \pm 2\rangle|0\rangle|2\rangle \\
	\left|\phi_{17,18}\right\rangle=|0 \pm 1\rangle|2\rangle|0\rangle\;\;\;\;\;\;\;\;\;\;\;\;\;\;\;\;\;\;\;\;\;\;\;\;
	$$
	\label{1}
\end{multline}
Notice that a two-qubit maximally entangled state is sufficient to distinguish the above set (\ref{1}) by the protocol given in \cite{Cohen2008}. Obviously, the resource state must be shared between Alice and Bob. The resource state must not be shared either between Bob and Charlie or between Alice and Charlie. It can be verified that the set (\ref{1}) is distinguishable only in $AB|C$ cut.\\

Whereas, the converse of the above arguments is not true. i.e., there exists sets of states which are distinguishable in a particular bipartition, say, $AB|C$ but cannot be distinguished by sharing a two qubit maximally entangled state between $A$ and $B$. Consider another example here.
\begin{multline}
	$$
	\;\;\;\;\;\;\left|\phi_{1,2}\right\rangle=|0\pm1\rangle|0\rangle|1\rangle,\;\;\; \left|\phi_{3,4}\right\rangle=|1\rangle|0\pm 1\rangle|0\rangle\\
	\left|\phi_{5,6}\right\rangle=|0\rangle|1\rangle|0 \pm 1\rangle,\;\;\;  \left|\phi_{7,8}\right\rangle=|2\rangle|1\rangle|1 \pm 2\rangle \\
	\left|\phi_{9,10}\right\rangle=|1\rangle|1 \pm 2\rangle|2\rangle,\;\;\;  \left|\phi_{11,12}\right\rangle=|1 \pm 2\rangle|2\rangle|1\rangle \\
	\left|\phi_{13,14}\right\rangle=|1 \pm 2\rangle|0\rangle|2\rangle,\;\;
	\left|\phi_{15,16}\right\rangle=|2\rangle|0\rangle|0 \pm 1\rangle\\
	\left|\phi_{17,18}\right\rangle=|0\rangle|2\rangle|1\pm2\rangle\;\;\;\;\;\;\;\;\;\;\;\;\;\;\;\;\;\;\;\;\;\;\;
	$$
	\label{4}
\end{multline}
the set of states (\ref{4}) is distinguishable in $AB|C$ cut, but indistinguishable by using a two-qubit maximally entangled state shared between Alice and Bob.\\

Local irreducibility is stronger than nonlocality \cite{Halder2019strong nonlocality}, that is a locally irreducible set is locally
indistinguishable but not vice versa. The above examples are the sets which are locally irreducible (using by the technique described in \cite{bhunia2020}) in tripartition and distinguishable at least in one bipartition.

\subsection{Construction of genuinely nonlocal sets with lesser cardinality}
\label{A2}
Such a genuinely nonlocal product sets (GNPS) is not locally reducible when all the parties are in separate locations. However, when two of the parties came together it is not possible to identify all the states through nontrivial OPM. The GNPBs constructed in \cite{Rout2019} contains a large number of states about their respective dimensions also the generalization is not proper with the higher dimensional cases. So it is further an issue to find genuinely nonlocal product sets (GNPS) which contains lesser number of states about their dimensions and generalized properly for higher dimensional subsystems also.\\\\

{\bfseries Proposition 2}. In $3 \otimes 3 \otimes 3$, the 15 states
\begin{multline}
	$$
	\;\;\;\;\;\;\;\left|\phi_{1,2}\right\rangle=|0 \pm 1\rangle|0\rangle|1\rangle, \;\; \left|\phi_{3,4}\right\rangle=|1\rangle|0 \pm 1\rangle|0\rangle \\
	\left|\phi_{5,6}\right\rangle=|0\rangle|1\rangle|0 \pm 1\rangle,\;\;  \left|\phi_{7,8}\right\rangle=|2\rangle|1\rangle|1 \pm 2\rangle \\
	\left|\phi_{9,10}\right\rangle=|0\rangle|0 \pm 2\rangle|0\rangle,\;\;  \left|\phi_{11,12}\right\rangle=|1 \pm 2\rangle|2\rangle|0\rangle \\
	\left|\phi_{13}\right\rangle=|0\rangle|2\rangle|1+2\rangle, \;\; \left|\phi_{14}\right\rangle=|0\rangle|0+1\rangle|2\rangle \\
	\left|\phi_{15}\right\rangle=|2\rangle|0\rangle|1+2\rangle\;\;\;\;\;\;\;\;\;\;\;\;\;\;\;\;\;\;\;\;
	$$
	\label{B}
\end{multline}
are locally irreducible in tripartition and also locally reducible but not completely distinguishable in every bipartite cut.\\\\

Proof: When the three parties are spatially separated, it is not
possible to eliminate any state by OPM from the above set and
hence the set is locally indistinguishable. This can be easily
proved by the technique described in \cite{Halder2019strong nonlocality,bhunia2020}. We are yet to
prove local indistinguishability of the above set across every
bipartition.\\
The set of states in $AB|C$ cut can be rewritten as,
\begin{multline}
	$$
	\;\;\;\;\;\;\;\;\;\;\;\left|\phi_{1,2}\right\rangle=|0 \pm 3\rangle|1\rangle,\;\;  \left|\phi_{3,4}\right\rangle=|3 \pm 4\rangle|0\rangle \\
	\left|\phi_{5,6}\right\rangle=|1\rangle|0 \pm 1\rangle,\;\;  \left|\phi_{7,8}\right\rangle=|7\rangle|1 \pm 2\rangle \\
	\left|\phi_{9,10}\right\rangle=|0 \pm 2\rangle|0\rangle, \;\; \left|\phi_{11,12}\right\rangle=|5 \pm 8\rangle|0\rangle \\
	\left|\phi_{13}\right\rangle=|2\rangle|1+2\rangle,\;\; \left|\phi_{14}\right\rangle=|0+1\rangle|2\rangle \\
	\left|\phi_{15}\right\rangle=|6\rangle|1+2\rangle\;\;\;\;\;\;\;\;\;\;\;\;\;\;\;\;\;\;\;\;\;
	$$
\end{multline}
\begin{figure}
	\centering
	\includegraphics[width=3.5in, height=2.1in]{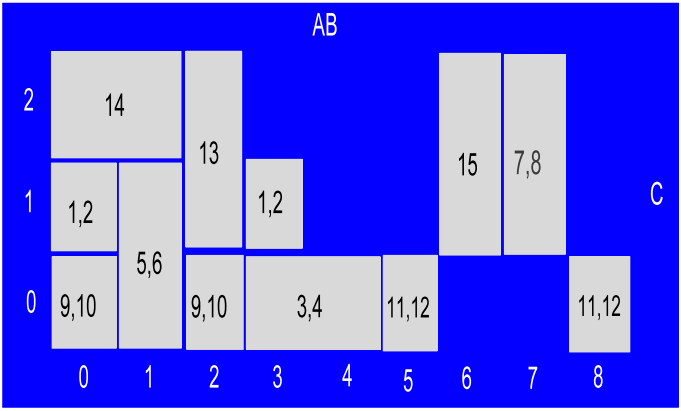}
	\caption{Structure of the set of states (\ref{B}) in the AB|C cut. The gray regions with indices $"i"$ represent the states $|\psi_i\rangle$.}
\label{F1}
\end{figure}
It is clear that Charlie ($C$) cannot start a nontrivial nondisturbing measurement. But $AB$ can start a nontrivial measurement\cite{bhunia2020}. If $AB$ start a protocol then only five states can be be eliminated $\left|\phi_{7,8}\right\rangle$, $\left|\phi_{11,12}\right\rangle$, $\left|\phi_{15}\right\rangle$. After that Charlie can distinguish those by particular choices of projectors. But the remaining ten states are still locally indistinguishable.\\

Now the set of states in $A|BC$ cut can be rewritten as,
\begin{multline}
$$
	\;\;\;\;\;\;\;\;\;\;\;\;\left|\phi_{1,2}\right\rangle=|0 \pm 1\rangle|1\rangle, \;\; \left|\phi_{3,4}\right\rangle=|1\rangle|0 \pm 3\rangle \\
	\left|\phi_{5,6}\right\rangle=|0\rangle|3 \pm 4\rangle, \;\; \left|\phi_{7,8}\right\rangle=|2\rangle|4 \pm 5\rangle \\
	\left|\phi_{9,10}\right\rangle=|0\rangle|0 \pm 6\rangle, \;\; \left|\phi_{11,12}\right\rangle=|1 \pm 2\rangle|6\rangle \\
	\left|\phi_{13}\right\rangle=|0\rangle|7+8\rangle,\;\; \left|\phi_{14}\right\rangle=|0\rangle|2+5\rangle \\
	\left|\phi_{15}\right\rangle=|2\rangle|1+2\rangle\;\;\;\;\;\;\;\;\;\;\;\;\;\;\;\;\;\;\;\;\;
	$$
\end{multline}
and the set of states in $AC|B$ cut can be rewritten as,
\begin{multline}
	$$
	\;\;\;\;\;\;\;\;\;\;\;\;\left|\phi_{1,2}\right\rangle=|1 \pm 4\rangle|0\rangle, \left|\phi_{3,4}\right\rangle=|3\rangle|0 \pm 1\rangle \\
	\left|\phi_{5,6}\right\rangle=|0 \pm 1\rangle|1\rangle, \left|\phi_{7,8}\right\rangle=|7 \pm 8\rangle|1\rangle \\
	\left|\phi_{9,10}\right\rangle=|0\rangle|0 \pm 2\rangle, \left|\phi_{11,12}\right\rangle=|3 \pm 6\rangle|2\rangle \\
	\left|\phi_{13}\right\rangle=|1+2\rangle|2\rangle, \left|\phi_{14}\right\rangle=|2\rangle|0+1\rangle \\
	\left|\phi_{15}\right\rangle=|7+8\rangle|0\rangle
	\;\;\;\;\;\;\;\;\;\;\;\;\;\;\;\;\;\;\;\;
	$$
\end{multline}\\

Using the similar argument which we have described for $AB|C$ cut, it can be seen that the set of states (\ref{B}) is locally reducible but not completely distinguishable in other two bipartition $A|BC$ and $AC|B$. \(\blacksquare \)\\\\

The set of states described above (\ref{B}) contains very less number of states to be genuinely nonlocal compared with its dimension. It is only because of the particular choice of twisted states as shown in FIG.~\ref{F1}. 
Next, we generalize nontrivially the class of states (\ref{B}) for higher dimensional subsystems.\\\\

{\bfseries Proposition 3}. In \({\mathbb {C}}^{d}\otimes {\mathbb {C}}^{d}\otimes {\mathbb {C}}^{d}\), where d is
odd, the set of \(\frac{15}{2}(d-1)\) orthogonal product states\\\\
\(|\phi_{1,2}\rangle =|0\pm1\rangle |0\rangle |\frac{d-1}{2}\rangle \).\\
\(|\phi_{3,4}\rangle =|\frac{d-1}{2}\rangle |0\pm1\rangle |0\rangle \).\\
\(|\phi_{5,6}\rangle =|0\rangle |\frac{d-1}{2}\rangle |0\pm1\rangle \).\\
\(|\phi_{5+i,6+i}\rangle =|i\pm \overline{i+1}\rangle |0\rangle |\frac{d-1}{2}\rangle \), \(i=2,4,6,...,(d-3).\)\\
\(|\phi_{d+2+i,d+3+i}\rangle =|\frac{d-1}{2}\rangle |i\pm \overline{i+1}\rangle |0\rangle \), \(i=2,4,6,...,(d-3).\)\\
\(|\phi_{2d-1+i,2d+i}\rangle =|0\rangle |\frac{d-1}{2}\rangle |i\pm \overline{i+1}\rangle \), \(i=2,4,6,...,(d-3).\)\\
\(|\phi_{3d-3+i,3d-2+i}\rangle =|d-1\rangle |\frac{d-1}{2}\rangle |i\pm \overline{i+1}\rangle \), \(i=1,3,5,...,(d-2).\)\\
\(|\phi_{4d-3+2i,4d-2+2i}\rangle =|0\rangle |i\pm \overline{d-i-1}\rangle|0\rangle\), \(i=0,1,2,...,(\frac{d-1}{2}-1).\)\\
\(|\phi_{5d-5+i,5d-4+i}\rangle =|i\pm \overline{i+1}\rangle |d-1\rangle |0\rangle \), \(i=1,3,5,...,(d-2).\)\\
\(|\phi_{6d-6+\frac{i+1}{2}}\rangle =|0\rangle |d-1\rangle |i\pm \overline{i+1}\rangle \), \(i=1,3,5,...,(d-2).\)\\
\(|\phi_{\frac{13d-13}{2}+\frac{i+2}{2}}\rangle =|0\rangle |i\pm \overline{i+1}\rangle |d-1\rangle \), \(i=0,2,4,...,(d-3).\)\\
\(|\phi_{7d-7+\frac{i+1}{2}}\rangle =|d-1\rangle |0\rangle |i\pm \overline{i+1}\rangle \), \(i=1,3,5,...,(d-2).\)\\\\
are locally irreducible in tripartition and also locally reducible but not completely distinguishable in every bipartite cut.\\\\

$Proof:$ Proof is similar as discussed above in Proposition 2. Firstly, we can show that the set of states is locally irreducible if all the parties are in separate locations by using the method we have discussed in \cite{bhunia2020}. Next, similar to Proposition 2, we observe that the set of states is not completely distinguishable if any two parties come together at the same location. \(\blacksquare \)\\\\

{\bfseries Proposition 4}. In  ${\mathbb{C}}^{d}\bigotimes{\mathbb{C}}^{d}\bigotimes{\mathbb{C}}^{d}$, where $d\geq4$ and $d$ is even, the set of $(18d-13)$ orthogonal product states\\\\
$|\phi_{i+1}^\pm\rangle=|i\pm\overline{i+1}\rangle|i\rangle|d\rangle$, $i=0,1,2,...,d-1.$\\
$|\phi_{d+1+i}^\pm\rangle=|d\rangle|i\pm\overline{i+1}\rangle|i\rangle$, $i=0,1,2,...,d-1.$\\
$|\phi_{2d+1+i}^\pm\rangle=|i\rangle|d\rangle|i\pm\overline{i+1}\rangle$, $i=0,1,2,...,d-1.$\\
$|\phi_{2d+1+i}^\pm\rangle=|i\pm\overline{i+1}\rangle|d-1\rangle|i+1\rangle$,$i=d,d+1,d+2,...,2d-2.$\\
$|\phi_{3d+i}^\pm\rangle=|i+1\rangle|i\pm\overline{i+1}\rangle|d-1\rangle$,  $i=d,d+1,d+2,...,2d-2.$\\
$|\phi_{4d-1+i}^\pm\rangle=|d-1\rangle|i+1\rangle|i\pm\overline{i+1}\rangle$,$i=d,d+1,d+2,...,2d-2.$\\
$|\phi_{6d-2+\frac{i-1}{2}}\rangle=|\overline{d-1}+d\rangle|i\rangle|d\rangle$, $i=1,3,5,...,d-3.$\\
$|\phi_{6d-1+\frac{d-4}{2}+\frac{i-1}{2}}\rangle=|d\rangle|\overline{d-1}+d\rangle|i\rangle$, $i=1,3,5,...,d-3.$\\
$|\phi_{7d-4+\frac{i-1}{2}}\rangle=|i\rangle|d\rangle|\overline{d-1}+d\rangle$, $i=1,3,5,...,d-3.$\\
$|\psi_{7d-3+\frac{d-4}{2}+\frac{i}{2}}\rangle=|\overline{d-2}+\overline{d-1}\rangle|i\rangle|d\rangle$, $i=0,2,4,...,d-4.$\\
$|\psi_{8d-6+\frac{i}{2}}\rangle=|d\rangle|\overline{d-2}+\overline{d-1}\rangle|i\rangle$, $i=0,2,4,...,d-4.$\\
$|\psi_{8d-5+\frac{d-4}{2}+\frac{i}{2}}\rangle=|i\rangle|d\rangle|\overline{d-2}+\overline{d-1}\rangle$, $i=0,2,4,...,d-4.$\\
$|\psi_{9d-9+\frac{i-d}{2}}\rangle=|\overline{d-1}+d\rangle|d-1\rangle|i\rangle$,$i=d+2,d+4,...,2d-2.$\\
$|\psi_{9d-10+\frac{i}{2}}\rangle=|i\rangle|\overline{d-1}+d\rangle|d-1\rangle$, $i=d+2,d+4,...,2d-2.$\\
$|\psi_{10d-11+\frac{i-d}{2}}\rangle=|d-1\rangle|i\rangle|\overline{d-1}+d\rangle$, $i=d+2,d+4,...,2d-2.$\\
$|\psi_{10d-11+\frac{i-3}{2}}\rangle=|d+\overline{d+1}\rangle|d-1\rangle|i\rangle$, $i=d+3,d+5,...,2d-1.$\\
$|\psi_{11d-13+\frac{i-d-1}{2}}\rangle=|i\rangle|d+\overline{d+1}\rangle|d-1\rangle$, $i=d+3,d+5,...,2d-1.$\\ $|\psi_{11d-13+\frac{i-3}{2}}\rangle=|d-1\rangle|i\rangle|d+\overline{d+1}\rangle$, $i=d+3,d+5,...,2d-1.$\\
$|\psi_{12d-14}\rangle=|\frac{d}{2}-1\rangle|\overline{\frac{d}{2}-1}+\frac{d}{2}\rangle|\frac{d}{2}\rangle$\\
$|\psi_{12d-13}\rangle=|\frac{d}{2}\rangle|\frac{d}{2}-1\rangle|\overline{\frac{d}{2}-1}+\frac{d}{2}\rangle$\\
$|\psi_{12d-12}\rangle=|\overline{\frac{d}{2}-1}+\frac{d}{2}\rangle|\frac{d}{2}\rangle|\frac{d}{2}-1\rangle$\\
$|\psi_{12d-11}\rangle=|\frac{d}{2}\rangle|\frac{d}{2}+\overline{\frac{d}{2}+1}\rangle|\frac{d}{2}\rangle$\\
$|\psi_{12d-10}\rangle=|\frac{d}{2}+1\rangle|\frac{d}{2}\rangle|\frac{d}{2}+\overline{\frac{d}{2}+1}\rangle$\\\\
are locally irreducible in tripartition and also locally reducible but not completely distinguishable in every bipartite cut.\\

$Proof:$ Proof is similar as discussed above in Proposition 2.\\

In \cite{Rout2019} authors construct a complete basis(Type-II(a) GNPB) in $3\otimes3$, which has same characteristics in the sense of genuine nonlocality as (\ref{B}). Also in \cite{Rout2019}, authors derive a protocol to distinguish them and showed that two-copies of two qubit maximally entangled state is sufficient as a resource to distinguish the class of states. Obviously the protocol used in \cite{Rout2019} is resource efficient. In \cite{Rout2020} provides a relative ordering relation of resource state for particular discrimination task. Further in \cite{Rout2020}, it was mentioned that the particular set(Type-II(a) GNPB) cannot be distinguished by using a GHZ state as a resource, i.e., for the particular discrimination task using two-copies of two qubit Bell state gives more advantage than a three qubit GHZ state. Here we construct a set of states (\ref{B}) contains minimum number of states and has the same characteristics in the sense of genuine nonlocality. After that we provide a discriminating protocol by using a three qubit GHZ state as a resource. The entanglement assisted discrimination protocol is provided in the next section. As the set of states are locally indistinguishable in every bipartition, then the set of states cannot be distinguished by using a two qubit maximally entangled state shared between any two parties as a resource. Therefore for the discrimination task of (\ref{B}), a three qubit GHZ state provides us advantage.

\subsection{Construction of strong nonlocal sets with lesser cardinality}
\label{A3}
Such a nonlocal set is locally irreducible even if any two parties come together. In \cite{Halder2019strong nonlocality} Halder et al., construct a genuine nonlocal product basis in $3\otimes3\otimes3$ which is locally irreducible in every bipartition and left an open question whether there exists an incomplete orthogonal product basis which is locally irreducible in every bipartition or not. Afterthat Rout et al. in\cite{Rout2019} provides a resouece efficient discrimination protocol  to distinguish that product basis(Type-II(b) GNPB). They showed that (1+$\log_23$) ebits is sufficient to discriminate this product basis. Recently Yuan et al., \cite{Yuan2020} construct an incomplete orthogonal basis in $3\otimes3\otimes3$ which is locally irreducible in every bipartition. Also they generalize the result for higher dimensional subsystems\cite{Yuan2020}.\\\\

{\bfseries (Previous result \cite{Yuan2020})}. The set of 24 states in $3 \otimes 3 \otimes 3$,
\begin{multline}
	$$
	\;\;\;\;\;\;\left|\phi_{1,2}\right\rangle=|0\rangle|1\rangle|0\pm1\rangle,\;\;\;\left|\phi_{3,4}\right\rangle=|0\pm1\rangle|0\rangle|1\rangle\\
	\left|\phi_{5,6}\right\rangle=|1\rangle|0 \pm 1\rangle|0\rangle,\;\;\;\;\left|\phi_{7,8}\right\rangle=|0\rangle|2\rangle|0\pm2\rangle\;\;\\
	\left|\phi_{9,10}\right\rangle=|0\pm2\rangle|0\rangle|2\rangle,\;\;\;\;\;\left|\phi_{11,12}\right\rangle=|2\rangle|0\pm2\rangle|0\rangle\\
	\left|\phi_{13,14}\right\rangle=|1\rangle|2\rangle|0 \pm1\rangle,\;\;\;\;\left|\phi_{15,16}\right\rangle=|0\pm1\rangle|1\rangle|2\rangle\\
	\left|\phi_{17,18}\right\rangle=|2\rangle|0\pm 1\rangle|1\rangle,\; \quad\left|\phi_{19,20}\right\rangle=|2\rangle|1\rangle|0\pm2\rangle\\
	\;\;\;\;\left|\phi_{21,22}\right\rangle=|0\pm2\rangle|2\rangle|1\rangle,\;\quad\left|\phi_{23,24}\right\rangle=|1\rangle|0\pm2\rangle|2\rangle\;\;
	$$
	\label{C}
\end{multline}
 are locally irreducible in tripartition and also locally irreducible in every bipartite cut.\\\\
 
Clearly, the class of states (\ref{C}) is the stronger form of nonlocality from the perspective of local elimination. In \cite{Yuan2020} authors described the complete proof for testing irreducibility in tripartition and also in every bipartition.
Interestingly it can be seen that the set of states (\ref{C}) is distinguishable by using (1+$\log_23$) ebits of entanglement as a resource. We derive an entanglement efficient discrimination protocol for distinguishing the above set of product states. The entanglement assisted discrimination protocol is provided in the next section.\\\\

All strong nonlocal sets \cite{Halder2019strong nonlocality,Yuan2020} described above have party symmetric structure. It also to be noted that they consumes the same amount of entanglement as a resource for their respective discrimination tasks. Therefore an open question arise whether it is possible to construct an incomplete strong nonlocal set which has party asymmetry structure and for the discrimination task which consumes (1+$\log_23$) ebits of entanglement is not sufficient. The following proposition will answer the above question preciously.\\\\

{\bfseries Proposition 5}. In $3 \otimes 3 \otimes 3$, the 26 states
\begin{multline}
	$$
	\;\;\left|\phi_{1,2,3,4}\right\rangle=|0 \pm 2\rangle|0 \pm 1\rangle|2\rangle,\;\left|\phi_{5,6}\right\rangle=|1\rangle|0\rangle|0 \pm 2\rangle\\
	\left|\phi_{7,8}\right\rangle=|1\rangle|0 \pm 2\rangle|1\rangle,\;\left|\phi_{9,10,11,12}\right\rangle=|0 \pm 2\rangle|0\rangle|0 \pm 1\rangle\\
	\left|\phi_{13,14}\right\rangle=|1\rangle|1\rangle|1 \pm 2\rangle,\;\;\left|\phi_{15,16}\right\rangle=|1 \pm 2\rangle|2\rangle|2\rangle\\
	\left|\phi_{17,18,19,20}\right\rangle=|2\rangle|1 \pm 2\rangle|0 \pm1\rangle,\;\left|\phi_{21,22}\right\rangle=|0\rangle|2\rangle|0 \pm 2\rangle\\
	\left|\phi_{23,24}\right\rangle=|0\rangle|1 \pm 2\rangle|1\rangle,\; \quad\left|\phi_{25,26}\right\rangle=|0 \pm 1\rangle|1\rangle|0\rangle\;\;
	$$
	\label{A}
\end{multline}
are locally irreducible in tripartite cut and also locally irreducible in every bipartite cut.\\\\

\begin{figure}
	\centering
	\includegraphics[width=3.5in, height=2.1in]{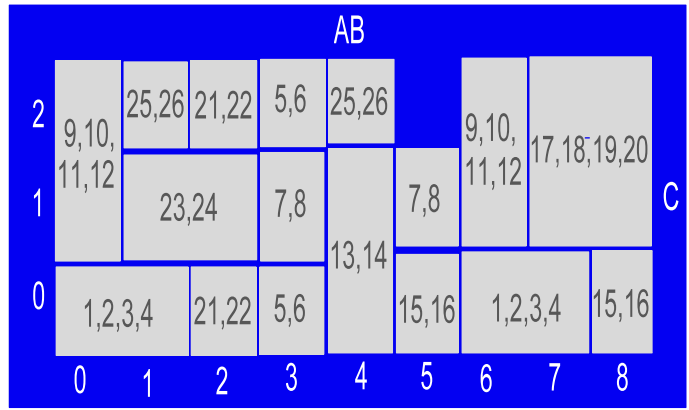}
	\caption{Structure of the set of states (\ref{A}) in the AB|C cut. The gray regions with indices $"i"$ represent the states $|\psi_i\rangle$.}
\label{F2}
\end{figure}

Proof: As the structure of the above set of states (\ref{A}) is asymmetric about all parties. Therefore, it is needed to show that any individual party cannot start a nontrivial and nondisturbing measurement, i.e., all measurements $E_{m}=M_{m}^{\dagger} M_{m}$ corresponding to each of the parties are proportional to the identity and as a consequence any one of them cannot eliminate any states.\\
 Suppose Alice starts with the nontrivial and nondisturbing measurement $E_{m}$. We write measurement $E_{m}$ in the $\{|0\rangle,|1\rangle,|2\rangle\}_{A}$ basis,
$$
E_{m}=\left[\begin{array}{lll}
	a_{00} & a_{01} & a_{02} \\
	a_{10} & a_{11} & a_{12} \\
	a_{20} & a_{21} & a_{22}
\end{array}\right]
$$
The postmeasurement states $\left\{M_{m} \otimes I_{B} \otimes I_{C}\left|\phi_{i}\right\rangle, i=\right.$
$1, \ldots, 26\}$ should be mutually orthogonal. We want to show that $E_{m}$ is proportional to the identity. Same for the other two parties also. The complete analysis of the proof is given in the Appendix A.\\
Now we need to show that the above set of states form a locally irreducible set in every bipartition. First we consider the set of states in $AB|C$ cut. Physically this means that the subsystems $A$ and $B$ are treated together as a nine-dimensional subsystem $AB$ on which joint measurements are now allowed. To reflect this, we rewrite the states (\ref{A}) as follows and depicted in FIG.~\ref{F2} :
\begin{multline}
	$$
	\;\;\left|\phi_{1}\right\rangle=|0+6+1+7\rangle|2\rangle,\;\left|\phi_{2}\right\rangle=|0+6-1-7\rangle|2\rangle\\
	\left|\phi_{3}\right\rangle=|0-6+1-7\rangle|2\rangle,\;\left|\phi_{4}\right\rangle=|0-6-1+7\rangle|2\rangle\\
	\left|\phi_{5,6}\right\rangle=|3\rangle|0 \pm 2\rangle,\;\left|\phi_{7,8}\right\rangle=|3 \pm 5\rangle|1\rangle\\
    \left|\phi_{9,10,11,12}\right\rangle=|0\pm6\rangle|0 \pm 1\rangle,\;\left|\phi_{13,14}\right\rangle=|4\rangle|1\pm2\rangle\\
	\left|\phi_{15,16}\right\rangle=|5\pm8\rangle|2\rangle,\;\left|\phi_{17,18,19,20}\right\rangle=|7 \pm 8\rangle|0\pm1\rangle\\
	\left|\phi_{21,22}\right\rangle=|2\rangle|0 \pm 2\rangle,\;\left|\phi_{23,24}\right\rangle=|1\pm 2\rangle|1\rangle\\
	\left|\phi_{25,26}\right\rangle=|1\pm4\rangle|0\rangle\;\;\;\;\;\;\;\;\;\;\;\;\;\;\;\;\;
	$$
	\label{13}
\end{multline}
Let $A B$ start with the nontrivial and nondisturbing measurement $E_{m}=M_{m}^{\dagger} M_{m} .$ We write measurement $E_{m}$ in the $\{|0\rangle,|1\rangle, \ldots,|8\rangle\}_{A B}$ basis, which corresponds to the states (\ref{13})\\
$E_{m}=\left[\begin{array}{lllllllll}a_{00} & a_{01} & a_{02} & a_{03} & a_{04} & a_{05} & a_{06} & a_{07} & a_{08} \\ a_{10} & a_{11} & a_{12} & a_{13} & a_{14} & a_{15} & a_{16} & a_{17} & a_{18} \\ a_{20} & a_{21} & a_{22} & a_{23} & a_{24} & a_{25} & a_{26} & a_{27} & a_{28} \\ a_{30} & a_{31} & a_{32} & a_{33} & a_{34} & a_{35} & a_{36} & a_{37} & a_{38} \\ a_{40} & a_{41} & a_{42} & a_{43} & a_{44} & a_{45} & a_{46} & a_{47} & a_{48} \\ a_{50} & a_{51} & a_{52} & a_{53} & a_{54} & a_{55} & a_{56} & a_{57} & a_{58} \\ a_{60} & a_{61} & a_{62} & a_{63} & a_{64} & a_{65} & a_{66} & a_{67} & a_{68} \\ a_{70} & a_{71} & a_{72} & a_{73} & a_{74} & a_{75} & a_{76} & a_{77} & a_{78} \\ a_{80} & a_{81} & a_{82} & a_{83} & a_{84} & a_{85} & a_{86} & a_{87} & a_{88}\end{array}\right] .$\\\\

The measurement must leave the postmeasurement states
mutually orthogonal. By choosing proper choice of pairs of vectors $\left\{\left|\phi_{i}\right\rangle,\left|\phi_{j}\right\rangle\right\}, i \neq j,$ we can find that all the off-diagonal matrix elements $a_{i j}, i \neq j$ must be zero if the orthogonality-preserving conditions $\left\langle\phi_{i}\left|E_{m} \otimes \mathbb{I}\right| \phi_{j}\right\rangle=0$ are to be satisfied. Similarly, we can find that the diagonal elements are all equal; i.e., overall we can show that $AB$ cannot start a nontrivial nondisturbing measurement. As we discuss previously that any single party cannot start a nontrivial nondisturbing measurement. Therefore the measurement corresponding $C$ must be trivial. So, (\ref{A}) is locally irreducible in the $AB|C$ cut. Same things will happen for the other two bipartitions $AC|B$ and $A|BC$. See Appendix A for the complete analysis. \(\blacksquare \)\\\\

As discussed above (1+$\log_23$) ebits of entanglement is sufficient to distinguish the incomplete strong nonlocal set \cite{Yuan2020}. While this amount of entanglement resource is not sufficient for the discrimination task of (\ref{A}). See Appendix C for this conjecture. From the best of our knowledge it is also been checked that three copies of two qubit Bell states or two copies of three qubit GHZ states is not sufficient for this discrimination task.
Now in the next section we study entanglement-assisted discrimination
protocols for the above classes of states.\\\\

\section{ENTANGLEMENT-ASSISTED DISCRIMINATION}
\label{A4}
In this section an entanglement-assisted local protocol is constructed to discriminate the states of the nonlocal completable set S as given in the previous section. Note that the normalization constants do not play any key role in the discrimination protocol. So, these constants are ignored for
simplicity in this entire section. As mentioned earlier, the mathematical structure of LOCC is still not clear and hence it is difficult to prove local distinguishability of a given set unless one constructs an explicit local protocol. This is why construction of a local protocol is so important.\\\\

{\bfseries Proposition 6}. One copy of three-qubit GHZ is sufficient to distinguish the set of states (\ref{B}).\\
$Proof:$ First, we assume that a GHZ state shared between three parties Alice, Bob and Charlie be $|\psi\rangle_{abc}$. Then the initial states shared among them is $$\left|\phi\right\rangle_{ABC}\otimes\left|\psi\right\rangle_{abc}$$
where $\left|\phi\right\rangle$ is one of the state from (\ref{B}).\\

Step $1 .$ Bob performs a measurement
$$
\begin{aligned}
	\mathcal{B} & \equiv\left\{M:=\mathbb{P}\left[(|0\rangle,|2\rangle)_{B} ;|0\rangle_{b}\right]+\mathbb{P}\left[|1\rangle_{B} ;|1\rangle_{b}\right]\right.\\
	\bar{M} &:=\mathbb{I}-M\}
\end{aligned}
$$
Suppose the outcomes corresponding to $M$ clicks. The resulting post-measurement states are therefore,\\
\begin{multline*}
	$$
	\left|\phi_{1,2}\right\rangle \rightarrow\left|0 \pm 1\rangle_{A}|0\rangle_{B}|1\rangle_{C}|000\rangle_{abc}\right.,\\
	\left|\phi_{3,4}\right\rangle \rightarrow\left|1\rangle_{A}|0\rangle_{B}|0\rangle_{C}|000\rangle_{abc}\right.\pm\left|1\rangle_{A}|1\rangle_{B}|0\rangle_{C}|111\rangle_{abc}\right.,\\
	\left|\phi_{5,6}\right\rangle \rightarrow\left|0\rangle_{A}|1\rangle_{B}|0\pm1\rangle_{C}|111\rangle_{abc}\right.,\;\;\;\;\;\;\;\;\;\;\;\;\;\;\;\;\;\;\;\;\;\;\;\;\;\;\;\;\;\;\;\\
	\left|\phi_{7,8}\right\rangle \rightarrow\left|2\rangle_{A}|1\rangle_{B}|1\pm2\rangle_{C}|111\rangle_{abc}\right.,\;\;\;\;\;\;\;\;\;\;\;\;\;\;\;\;\;\;\;\;\;\;\;\;\;\;\;\;\;\;\\
	\left|\phi_{9,10}\right\rangle \rightarrow\left|0\rangle_{A}|0\pm2\rangle_{B}|0\rangle_{C}|000\rangle_{abc}\right.,\;\;\;\;\;\;\;\;\;\;\;\;\;\;\;\;\;\;\;\;\;\;\;\;\;\;\;\;\;\\
	\left|\phi_{11,12}\right\rangle \rightarrow\left|1\pm2\rangle_{A}|2\rangle_{B}|0\rangle_{C}|000\rangle_{abc}\right.,\;\;\;\;\;\;\;\;\;\;\;\;\;\;\;\;\;\;\;\;\;\;\;\;\;\;\;\;\\
	\left|\phi_{13}\right\rangle \rightarrow\left|0\rangle_{A}|2\rangle_{B}|1+2\rangle_{C}|000\rangle_{abc}\right.,\;\;\;\;\;\;\;\;\;\;\;\;\;\;\;\;\;\;\;\;\;\;\;\;\;\;\;\;\;\;\;\;\\
	\left|\phi_{14}\right\rangle \rightarrow\left|0\rangle_{A}|0\rangle_{B}|2\rangle_{C}|000\rangle_{abc}\right.+\left|0\rangle_{A}|1\rangle_{B}|2\rangle_{C}|111\rangle_{abc}\right.,\\
	\left|\phi_{15}\right\rangle \rightarrow\left|2\rangle_{A}|0\rangle_{B}|1+2\rangle_{C}|000\rangle_{abc}\right..\;\;\;\;\;\;\;\;\;\;\;\;\;\;\;\;\;\;\;\;\;\;\;\;\;\;\;\;\;\;\;\;\\
	$$
\end{multline*}
Step $2 .$ Alice performs a measurement
$$
\begin{aligned}
	\mathcal{A} & \equiv\left\{N:=\mathbb{P}\left[|2\rangle_{A} ;|1\rangle_{a}\right]\right.,\bar{N}:=\mathbb{I}-N\}
\end{aligned}
$$
If the outcomes corresponding to $N$ clicks. The resulting post measurement states are therefore $\left|2\rangle_{A}|1\rangle_{B}|1\pm2\rangle_{C}|111\rangle_{abc}\right.$, which can be easily distinguished by Charlie by projecting onto $|1\pm2\rangle_{C}$. If the outcome corresponding to $\bar{N}$ clicks, then the remaining 13 states are isolated. \\\\
Step $3 .$ Charlie performs a measurement
$$
\begin{aligned}
	\mathcal{C} & \equiv\left\{Q:=\mathbb{P}\left[|0\rangle_{C} ;|0\rangle_{c}\right]+\mathbb{P}\left[(|0\rangle,|1\rangle)_{C} ;|1\rangle_{c}\right]\right.\\
	\bar{Q} &:=\mathbb{I}-Q\}
\end{aligned}
$$
If the outcomes corresponding to $Q$ clicks. The resulting post measurement states are therefore $\left|1\rangle_{A}|0\rangle_{B}|0\rangle_{C}|000\rangle_{abc}\right.\pm\left|1\rangle_{A}|1\rangle_{B}|0\rangle_{C}|111\rangle_{abc}\right.$, $\left|0\rangle_{A}|1\rangle_{B}|0\pm1\rangle_{C}|111\rangle_{abc}\right.$, $\left|0\rangle_{A}|0\pm2\rangle_{B}|0\rangle_{C}|000\rangle_{abc}\right.$ and $\left|1\pm2\rangle_{A}|2\rangle_{B}|0\rangle_{C}|000\rangle_{abc}\right.$. Now Alice makes three outcome projective measurement $A_{1}=|0\rangle_{A}\langle0|\otimes|1\rangle_{a}\langle1|$, $A_{2}=|0\rangle_{A}\langle0|\otimes|0\rangle_{a}\langle0|$, $A_{3}=\mathbb{I}-A_{1}-A_{2}$. If $A_{1}$ clicks, it isolates two states $\left|0\rangle_{A}|1\rangle_{B}|0\pm1\rangle_{C}|111\rangle_{abc}\right.$, which can be easily distinguished by Charlie by projecting onto $|0\pm1\rangle_{C}$.  If $A_{2}$ clicks, it isolates two states $\left|0\rangle_{A}|0\pm2\rangle_{B}|0\rangle_{C}|000\rangle_{abc}\right.$, which can be easily distinguished by Bob by projecting onto $|0\pm2\rangle_{B}$. If the outcome $A_{3}$ clicks, it isolates four states $\left|1\rangle_{A}|0\rangle_{B}|0\rangle_{C}|000\rangle_{abc}\right.\pm\left|1\rangle_{A}|1\rangle_{B}|0\rangle_{C}|111\rangle_{abc}\right.$ and $\left|1\pm2\rangle_{A}|2\rangle_{B}|0\rangle_{C}|000\rangle_{abc}\right.$. Next, Bob makes two outcome projective measurement $B_{1}=|2\rangle_{B}\langle2|\otimes|0\rangle_{b}\langle0|$, $B_{2}=\mathbb{I}-B_{1}$. Now, if $B_{1}$ clicks, it isolates two states $\left|1\pm2\rangle_{A}|2\rangle_{B}|0\rangle_{C}|000\rangle_{abc}\right.$, which can be easily distinguished by Alice by projecting onto $|1\pm2\rangle_{A}$. If the outcome $B_{2}$ clicks, it isolates two states $\left|1\rangle_{A}|0\rangle_{B}|0\rangle_{C}|000\rangle_{abc}\right.\pm\left|1\rangle_{A}|1\rangle_{B}|0\rangle_{C}|111\rangle_{abc}\right.$. We know from Walgate et.al. result that any two pure bipartite/multipartite orthogonal states can be perfectly distinguishable by LOCC. Hence those two states are perfectly distinguished.\\\\

Next if the outcome $\bar{Q}$ clicks, the post measurement states are therefore, $\left|0\pm1\rangle_{A}|0\rangle_{B}|1\rangle_{C}|000\rangle_{abc}\right.$, $\left|0\rangle_{A}|2\rangle_{B}|1+2\rangle_{C}|000\rangle_{abc}\right.$, $\left|0\rangle_{A}|0\rangle_{B}|2\rangle_{C}|000\rangle_{abc}\right.+\left|0\rangle_{A}|1\rangle_{B}|2\rangle_{C}|111\rangle_{abc}\right.$ and $\left|2\rangle_{A}|0\rangle_{B}|1+2\rangle_{C}|000\rangle_{abc}\right.$. Bob makes then two outcome projective measurement $B_{1}=|2\rangle_{B}\langle2|\otimes|0\rangle_{b}\langle0|$, $B_{2}=\mathbb{I}-B_{1}$. If the outcome $B_{1}$ click, it isolates only one state $\left|0\rangle_{A}|2\rangle_{B}|1+2\rangle_{C}|000\rangle_{abc}\right.$. If the outcome $B_{2}$ clicks, it isolates the remaining four states $\left|0\pm1\rangle_{A}|0\rangle_{B}|1\rangle_{C}|000\rangle_{abc}\right.$, $\left|0\rangle_{A}|0\rangle_{B}|2\rangle_{C}|000\rangle_{abc}\right.+\left|0\rangle_{A}|1\rangle_{B}|2\rangle_{C}|111\rangle_{abc}\right.$ and $\left|2\rangle_{A}|0\rangle_{B}|1+2\rangle_{C}|000\rangle_{abc}\right.$. Now, Alice makes two outcome projective measurement $A_{1}=|0\rangle_{A}\langle0|\otimes|0\rangle_{a}\langle0|+|1\rangle_{A}\langle1|\otimes|0\rangle_{a}\langle0|+|0\rangle_{A}\langle0|\otimes|1\rangle_{a}\langle1|$, $A_{2}=\mathbb{I}-A_{1}$. If $A_{1}$ clicks, three states are isolated and can be distinguished by Charlie by projecting onto $|1\rangle_{C}\langle1|\otimes|0\rangle_{c}\langle0|$ and $|2\rangle_{C}\langle2|\otimes|0\rangle_{c}\langle0|+|2\rangle_{C}\langle2|\otimes|1\rangle_{c}\langle1|$ respectively. Also, if the outcome $A_{2}$ click, it indicates only one state  $\left|0\rangle_{A}|0\rangle_{B}|2\rangle_{C}|000\rangle_{abc}\right.+\left|0\rangle_{A}|1\rangle_{B}|2\rangle_{C}|111\rangle_{abc}\right.$. Hence the set is distinguished. \(\blacksquare \) \\\\

The next result looks at a scenario for entanglement-assisted discrimination protocol of the class of states in the higher dimensional case.\\\\

{\bfseries Proposition 7}. In \({\mathbb {C}}^{d}\bigotimes {\mathbb {C}}^{d}\bigotimes {\mathbb {C}}^{d}\), where $d$ is odd, a GHZ state shared between three parties is sufficient to perfectly distinguished the set of \(\frac{15}{2}(d-1)\)
orthogonal product states described in Proposition 3.\\

$Proof:$ The method is similar to the approach we have used previously for \(d=3\). First of all let us assume that a GHZ state shared between three parties Alice, Bob and Charlie be $|\psi\rangle_{abc}$. Therefore the initial state shared among them is $\left|\phi\right\rangle_{ABC}\otimes\left|\psi\right\rangle_{abc}$, where $\left|\phi\right\rangle$ is one of the state from Proposition 3. Afterthat Bob will perform a measurement
\begin{multline*}
	$$
	\mathcal{B}  \equiv\{M:=\mathbb{P}[(|0\rangle,..,|\frac{d-1}{2}-1\rangle,|\frac{d-1}{2}+1\rangle,..,\\
	|d-1\rangle)_{B} ;|0\rangle_{b}]+\mathbb{P}[|\frac{d-1}{2}\rangle_{B} ;|1\rangle_{b}]\\
	\bar{M} :=\mathbb{I}-M\}
	$$
\end{multline*}\\
Next Alice, Bob and Charlie will do a sequence of measurements to distinguish those states as we have done in
previous one. \(\blacksquare \)\\\\

 {\bfseries Proposition 8}. (1+$\log_23$) ebits of entanglement is sufficient to distinguish the above set of states (\ref{C}).\\\\
$Proof:$ Firstly let us assume that Bob will teleport his system to Alice by consuming $\log_2 3$ ebit of entanglement as a resource. Afterthat we will prove that a two qubit maximally entangled state shared between Alice and Charlie is sufficient to distinguish the set of state (\ref{C}). The complete proof is described in Appendix B. Therefore, in total,
(1+$\log_23$) ebit entanglement is consumed in this protocol,
which is strictly less than the amount consumed in the protocol
using teleportation in both arms. \(\blacksquare \)\\\\

Now it is interesting to know that whether the above amount of entanglement is sufficient for the discrimination task of the set (\ref{A}) or not. Using the two-qutrit maximally entangled state $(|00\rangle+$ $|11\rangle+|22\rangle) / \sqrt{3},$ any one party say Bob first teleports his subsystem to Alice. Entanglement consumed at this step amounts to $\log_2 3$ ebits. Now the system is in $9\otimes3$ dimension($AB|C$ cut). In \cite{Cohen2008} author describe that a $[m / 2] \otimes[m / 2]$ maximally entangled state is sufficient to perfectly distinguish the GenTiles2 UPB on $m \otimes n$ with $n \geq m$, for any dimensions $m, n$ in which it exists (excluding the case $m=3$ ). So it is quite curious to check whether a $9\otimes3$ nonlocal set can be distinguishable by a $2\otimes2$ maximally entangled state or not. But it can checked that the set is not completely distinguishable by a $2\otimes2$ maximally entangled state. The choice of all possible measurements described in Appendix C. Thus it can be concluded that (1+$\log_23$) ebit of entanglement is not sufficient to distinguish perfectly the set of states (\ref{A}).\\\\ 

{\bfseries Proposition 9}. The entanglement resource $\left|\phi^{+}(3)\right\rangle_{\mathcal{A B}}$ shared between $(Alice, Bob)$ and $\left|\phi^{+}(3)\right\rangle_{\mathcal{A C}}$ shared between $(Alice, Charlie)$  is sufficient for local discrimination
of the set of states (\ref{A})  where $\left|\phi^{+}(3)\right\rangle:=$
$(|00\rangle+|11\rangle+|22\rangle) / \sqrt{3} \in \mathbb{C}^{3} \otimes \mathbb{C}^{3}.$\\\\

Using the two-qutrit maximally entangled state $(|00\rangle+$ $|11\rangle+|22\rangle) / \sqrt{3},$ Bob first teleports his subsystem to Alice. Entanglement consumed at this step amounts to $\log_2 3$ ebits. Afterthat Charlie will teleport his system to Alice by using  two-qutrit maximally entangled state $(|00\rangle+$ $|11\rangle+|22\rangle) / \sqrt{3}$. Entanglement consumed at this step also amounts to $\log 3$ ebits. Then three parties jointly perform projective measurement to distinguish the above set (\ref{A}). Therefore, in total, $(\log_2 3+\log_2 3)=2\log_2 3$ ebits of entanglement is consumed in this protocol. \(\blacksquare \)\\\\
For the incomplete strong nonlocal set described in \cite{Yuan2020}, a two qubit Bell state shared between any bipartition is sufficient to distinguish the set. Therefore by Proposition 1, the set of states in \cite{Yuan2020} is distinguishable if the third party comes together with the other two parties who are already at the same location. Whereas the discrimination task of (\ref{A}) indicates that converse of Proposition 1 is not true. Because the set (\ref{A}) is distinguishable if all three parties are come together but cannot be distinguishable by using a two qubit Bell state shared between any bipartition.
\section{CONCLUSION AND OPEN PROBLEMS}
\label{A5}
	Local discrimination of quantum states has attracted much attention during the last twenty years. The local distinguishability of quantum states can be applied to design quantum protocols such as quantum cryptography. The construction of sets of locally indistinguishable multipartite orthogonal product states with only less number of members(>2) is more difficult than bipartite ones. Strong nonlocality without entanglement leads to the strongest part of the phenomenon quantum nonlocality without entanglement which is firstly proposed by Halder et al.,\cite{Halder2019strong nonlocality}. Therefore construction of strong nonlocal set with lesser number of states is an important issue of research in recent years. In this paper, firstly we have constructed party asymmetric genuine nonlocal sets in arbitrary large dimensional system $C^d\otimes C^d\otimes C^d$, where $d\geq3$. It is noted that the sets which are constructed include a considerable number of states. Also, we provide a discriminating protocol by using a three qubit GHZ state as a resource. This indicates strong advantage by using genuine entanglement as a resource. Secondly, we construct an incomplete party asymmetric strong nonlocal set in $C^3\otimes C^3\otimes C^3$ which needs much entanglement resource for its discrimination task and indicates strongness from before. Our results show the phenomenon of strong quantum nonlocality with entanglement. There are some interesting problems left. We don't know whether three-qubit GNPS exists or not. How to generalize the construction in $(d)^{\otimes n}$ for any $d\geq2$ and $n\geq4$? Next is to find the class of strong nonlocal set which we have constructed in Proposition 5, for arbitrary large dimensions. Also the question of optimality of the entangled resources used in our discrimination protocols remains open.
\section*{ACKNOWLEDGEMENTS}
The authors I. Chattopadhyay and D. Sarkar acknowledge the work as part of QUest initiatives by DST India. The author A. Bhunia acknowledges the support from UGC, India

\begin{widetext}
\section*{Appendix A: Proof of Proposition 5.}
To test the irreducibility in all tripartition of the set (\ref{A}) it is needed to show that any individual party cannot start a nontrivial and nondisturbing measurement. Suppose Alice starts with the nontrivial and nondisturbing measurement $E_{m}$. We write measurement $E_{m}$ in the $\{|0\rangle,|1\rangle,|2\rangle\}_{A}$ basis,
$$
E_{m}=\left[\begin{array}{lll}
	a_{00} & a_{01} & a_{02} \\
	a_{10} & a_{11} & a_{12} \\
	a_{20} & a_{21} & a_{22}
\end{array}\right]
$$
The postmeasurement states $\left\{M_{m} \otimes I_{B} \otimes I_{C}\left|\phi_{i}\right\rangle, i=\right.$
$1, \ldots, 26\}$ should be mutually orthogonal. We want to show that $E_{m}$ is proportional to the identity. Same for the other two parties also.\\
 Now choosing the appropriate states from the set (\ref{A}) it can be shown that all the measurement of either parties must proportional to identity. The next tables indicate that properly.\\\\
{\bfseries The measurement corresponding to the party $A$.}
$$
\begin{array}{|c|c|}
	\hline \text { States } & \text { Elements }\\
	\hline\left\langle\phi_{23}\left|E_{m} \otimes \mathbb{I} \otimes \mathbb{I}\right| \phi_{13}\right\rangle=0 & a_{01}=0 \\
	\left\langle\phi_{13}\left|E_{m} \otimes \mathbb{I} \otimes \mathbb{I}\right| \phi_{23}\right\rangle=0 & a_{10}=0 \\
	\hline\left\langle\phi_{1}\left|E_{m} \otimes \mathbb{I} \otimes \mathbb{I}\right| \phi_{5}\right\rangle=0 & a_{21}=0 \\
	\left\langle\phi_{5}\left|E_{m} \otimes \mathbb{I} \otimes \mathbb{I}\right| \phi_{1}\right\rangle=0 & a_{12}=0 \\
	\hline\left\langle\phi_{23}\left|E_{m} \otimes \mathbb{I} \otimes \mathbb{I}\right| \phi_{17}\right\rangle=0 & a_{02}=0 \\
	\left\langle\phi_{17}\left|E_{m} \otimes \mathbb{I} \otimes \mathbb{I}\right| \phi_{23}\right\rangle=0 & a_{20}=0 \\
	\hline\left\langle\phi_{25}\left|E_{m} \otimes \mathbb{I} \otimes \mathbb{I}\right| \phi_{26}\right\rangle=0 & a_{00}=a_{11} \\
	\left\langle\phi_{15}\left|E_{m} \otimes \mathbb{I} \otimes \mathbb{I}\right| \phi_{16}\right\rangle=0 & a_{11}=a_{22} \\
	\hline
\end{array}
$$
Therefore, $E_{m}$ is proportional to the identity. So Alice can do only trivial measurement. The similar is true for other two parties.\\\\
{\bfseries The measurement corresponding to the party $B$.}
$$
\begin{array}{|c|c|}
	\hline \text { States } & \text { Elements }\\
	\hline\left\langle\phi_{5}\left|\mathbb{I} \otimes E_{m} \otimes\mathbb{I}\right| \phi_{13}\right\rangle=0 & a_{01}=0 \\
	\left\langle\phi_{13}\left|\mathbb{I} \otimes E_{m} \otimes\mathbb{I}\right| \phi_{5}\right\rangle=0 & a_{10}=0 \\
	\hline\left\langle\phi_{21}\left|\mathbb{I} \otimes E_{m} \otimes\mathbb{I}\right| \phi_{25}\right\rangle=0 & a_{21}=0 \\
	\left\langle\phi_{25}\left|\mathbb{I} \otimes E_{m} \otimes\mathbb{I}\right| \phi_{21}\right\rangle=0 & a_{12}=0 \\
	\hline\left\langle\phi_{5}\left|\mathbb{I} \otimes E_{m} \otimes\mathbb{I}\right| \phi_{15}\right\rangle=0 & a_{02}=0 \\
	\left\langle\phi_{15}\left|\mathbb{I} \otimes E_{m} \otimes\mathbb{I}\right| \phi_{5}\right\rangle=0 & a_{20}=0 \\
	\hline\left\langle\phi_{1}\left|\mathbb{I} \otimes E_{m} \otimes\mathbb{I}\right| \phi_{2}\right\rangle=0 & a_{00}=a_{11} \\
	\left\langle\phi_{23}\left|\mathbb{I} \otimes E_{m} \otimes\mathbb{I}\right| \phi_{24}\right\rangle=0 & a_{11}=a_{22} \\
	\hline
\end{array}
$$
{\bfseries The measurement corresponding to the party $C$.}
$$
\begin{array}{|c|c|}
	\hline \text { States } & \text { Elements }\\
	\hline\left\langle\phi_{25}\left|\mathbb{I}\otimes\mathbb{I} \otimes E_{m}\right| \phi_{23}\right\rangle=0 & a_{01}=0 \\
	\left\langle\phi_{23}\left|\mathbb{I}\otimes\mathbb{I} \otimes E_{m}\right| \phi_{25}\right\rangle=0 & a_{10}=0 \\
	\hline\left\langle\phi_{15}\left|\mathbb{I}\otimes\mathbb{I} \otimes E_{m}\right| \phi_{7}\right\rangle=0 & a_{21}=0 \\
	\left\langle\phi_{7}\left|\mathbb{I}\otimes\mathbb{I} \otimes E_{m}\right| \phi_{15}\right\rangle=0 & a_{12}=0 \\
	\hline\left\langle\phi_{25}\left|\mathbb{I}\otimes\mathbb{I} \otimes E_{m}\right| \phi_{1}\right\rangle=0 & a_{02}=0 \\
	\left\langle\phi_{1}\left|\mathbb{I}\otimes\mathbb{I} \otimes E_{m}\right| \phi_{25}\right\rangle=0 & a_{20}=0 \\
	\hline\left\langle\phi_{9}\left|\mathbb{I}\otimes\mathbb{I} \otimes E_{m}\right| \phi_{10}\right\rangle=0 & a_{00}=a_{11} \\
	\left\langle\phi_{13}\left|\mathbb{I}\otimes\mathbb{I} \otimes E_{m} \right| \phi_{14}\right\rangle=0 & a_{11}=a_{22} \\
	\hline
\end{array}
$$
Therefore the set of states (\ref{A}) is irreducible in all the tripartition. Now we need to show that the set is irreducible in all bipartition also.

{\bfseries The set of states in $AB|C$ cut can be rewritten as,}
\begin{multline}
	$$
	\;\;\;\;\;\;\;\;\;\;\;\;\;\;\;\;\;\;\;\;\;\;\;\;\;\;\;\;\;\;\;\;\;\;\;\;\;\;\;\;\;\;\;\;\;\;\;\;\;\left|\phi_{1}\right\rangle=|0+6+1+7\rangle|2\rangle,\;\left|\phi_{2}\right\rangle=|0+6-1-7\rangle|2\rangle\\
	\left|\phi_{3}\right\rangle=|0-6+1-7\rangle|2\rangle,\;\left|\phi_{4}\right\rangle=|0-6-1+7\rangle|2\rangle\\
	\left|\phi_{5,6}\right\rangle=|3\rangle|0 \pm 2\rangle,\;\left|\phi_{7,8}\right\rangle=|3 \pm 5\rangle|1\rangle\\
	\left|\phi_{9,10,11,12}\right\rangle=|0\pm6\rangle|0 \pm 1\rangle,\;\left|\phi_{13,14}\right\rangle=|4\rangle|1\pm2\rangle\\
	\left|\phi_{15,16}\right\rangle=|5\pm8\rangle|2\rangle,\;\left|\phi_{17,18,19,20}\right\rangle=|7 \pm 8\rangle|0\pm1\rangle\\
	\left|\phi_{21,22}\right\rangle=|2\rangle|0 \pm 2\rangle,\;\left|\phi_{23,24}\right\rangle=|1\pm 2\rangle|1\rangle\\
	\left|\phi_{25,26}\right\rangle=|1\pm4\rangle|0\rangle\;\;\;\;\;\;\;\;\;\;\;\;\;\;\;\;\;\;\;\;\;\;\;\;\;\;\;\;\;\;\;\;\;\;\;\;\;\;\;\;\;\;\;\;\;\;\;\;\;\;\;\;\;\;\;\;\;\;\;\;\;\;\;\;\;
	$$
	\label{14}
\end{multline}
By choosing appropriate states from (\ref{14}) we can find out the $9\times9$ matrix elements corresponding to the measurement of $AB$,
$$
	\begin{array}{|c|c||c|c|}
		\hline \text { states } & \text { elements } & \text { states } & \text { elements } \\
		\hline \hline|5+8\rangle|2\rangle,|5-8\rangle|2\rangle & a_{55}=a_{88} & |7+8\rangle|0+1\rangle,|7-8\rangle|0+1\rangle & a_{77}=a_{88} \\
		\hline|3+5\rangle|1\rangle,|3-5\rangle|1\rangle & a_{33}=a_{55} & |1+2\rangle|1\rangle,|1-2\rangle|1\rangle & a_{11}=a_{22}\\
		\hline|1+4\rangle|0\rangle,|1-4\rangle|0\rangle & a_{11}=a_{44} & |0+6\rangle|0+1\rangle,|0-6\rangle|0+1\rangle & a_{00}=a_{66}\\
		\hline
\end{array}
$$
$$
\begin{array}{|c|c|c|}
	\hline \text { Serial Number } & \text { States } & \text { Elements } \\
	\hline \hline(1) & |4\rangle|1+2\rangle,|2\rangle|0+2\rangle & a_{42}=a_{24}=0\\
	\hline(2) & |4\rangle|1+2\rangle,|3\rangle|0+2\rangle & a_{43}=a_{34}=0\\
	\hline(3) & |4\rangle|1+2\rangle,|3+5\rangle|1\rangle & a_{45}=a_{54}=0\\
	\hline(4) & |4\rangle|1+2\rangle,|5+8\rangle|2\rangle & a_{48}=a_{84}=0 \\
	\hline(5) & |4\rangle|1+2\rangle,|7+8\rangle|0+1\rangle & a_{47}=a_{74}=0\\
	\hline(6) & |4\rangle|1+2\rangle,|1+2\rangle|1\rangle & a_{41}=a_{14}=0 \\
	\hline(7) & |4\rangle|1+2\rangle,|0\pm6\rangle|0+1\rangle & a_{40}=a_{04}=a_{46}=a_{64}=0\\
	\hline(8) & |2\rangle|0+2\rangle,|1+4\rangle|0\rangle & a_{21}=a_{12}=0 \\
	\hline(9) & |2\rangle|0+2\rangle,|0\pm6\rangle|0+1\rangle & a_{20}=a_{02}=a_{26}=a_{62}=0 \\
	\hline(10) & |2\rangle|0+2\rangle,|3\rangle|0+2\rangle & a_{23}=a_{32}=0\\
    \hline(11) & |2\rangle|0+2\rangle,|7\pm8\rangle|0+1\rangle & a_{27}=a_{72}=a_{28}=a_{82}=0 \\
    \hline(12) & |2\rangle|0+2\rangle,|5+8\rangle|2\rangle & a_{25}=a_{52}=0\\
    \hline(13) & |3\rangle|0+2\rangle,|0\pm6\rangle|0+1\rangle & a_{30}=a_{03}=a_{36}=a_{63}=0 \\
    \hline(14) & |3\rangle|0+2\rangle,|1+4\rangle|0\rangle & a_{31}=a_{13}=0\\
    \hline(15) & |3\rangle|0+2\rangle,|5\pm8\rangle|2\rangle & a_{35}=a_{53}=a_{38}=a_{83}=0 \\
    \hline(16) & |3\rangle|0+2\rangle,|7+8\rangle|0+1\rangle & a_{37}=a_{73}=0\\
	\hline(17) & |3+5\rangle|1\rangle,|0\pm6\rangle|0+1\rangle & a_{50}=a_{05}=a_{56}=a_{65}=0 \\
	\hline(18) & |3+5\rangle|1\rangle,|1+2\rangle|1\rangle & a_{51}=a_{15}=0\\
	\hline(19) & |3+5\rangle|1\rangle,|7\pm8\rangle|0+1\rangle & a_{57}=a_{75}=a_{58}=a_{85}=0 \\
	\hline(20) & |3+5\rangle|1\rangle,|1+2\rangle|1\rangle & a_{51}=a_{15}=0\\
	\hline(21) & |0\pm6\rangle|0+1\rangle,|1+2\rangle|1\rangle & a_{01}=a_{10}=a_{61}=a_{16}=0 \\
	\hline(22) & |3+5\rangle|1\rangle,|1+2\rangle|1\rangle & a_{51}=a_{15}=0\\
	\hline(23) & |7\pm8\rangle|0+1\rangle,|1+4\rangle|0\rangle & a_{17}=a_{71}=a_{18}=a_{81}=0 \\
	\hline(24) & |3+5\rangle|1\rangle,|1+2\rangle|1\rangle & a_{51}=a_{15}=0\\
	\hline
\end{array}
$$
the remaining elements are $$a_{06},a_{60},a_{07},a_{70},a_{08},a_{80},a_{67},a_{76},a_{68},a_{86},a_{78},a_{87}.$$
Now if we consider the states $|0+6+1+7\rangle|2\rangle,|0+6-1-7\rangle|2\rangle,|0-6+1-7\rangle|2\rangle,|0-6-1+7\rangle|2\rangle$ and the state $|5+8\rangle|2\rangle$, we can get the elements $a_{80}=a_{08}=a_{86}=a_{68}=a_{87}=a_{78}=0$. Next if we take $|0\pm6\rangle|0+1\rangle,|7+8\rangle|0+1\rangle$, we get the elements $a_{70}=a_{07}=a_{76}=a_{67}=0$. The remaining states are $a_{60},a_{06}$.\\
So, if we take the states $|0+6+1+7\rangle|2\rangle,|0+6-1-7\rangle|2\rangle$ and since The measurement must leave the postmeasurement states
mutually orthogonal. Therefore we have,\\
\[\begin{aligned}&\left\langle\phi_{i}\left|E_{m} \otimes \mathbb{I}\right| \phi_{j}\right\rangle=0
	\\&\quad \implies \langle 0+6+1+7|E_{m}|0+6-1-7\rangle \langle 2||2\rangle=0 \\&\quad \implies a_{00}+a_{06}-a_{01}-a_{07}+a_{60}+a_{66}-a_{61}-a_{67}+a_{10}+a_{16}-a_{11}-a_{17}+a_{70}+a_{76}-a_{71}-a_{77}=0 \\&\quad \implies
	a_{06}+a_{60}=0
	\\&\quad \implies
	2a_{06}=0
	\\&\quad \implies
	a_{06}=0
\end{aligned}\]
 [All the remaining terms of the equation becomes cancelled out by using the conditions $a_{ij}=a_{ji}=0, \;\;\;i,j\neq0,6$ \;\;\;and $a_{ii}=a_{jj} \;\;\;\;\forall i,j.$]\\
 Therefore we have $a_{06}=a_{60}=0$.\\
 Thus,  $AB$ cannot start a nontrivial nondisturbing measurement. Also we show previously that the measurement corresponding to $C$ must be trivial. Hence the above set of states is locally irreducible in $AB|C$ bipartition.

{\bfseries The set of states in $A|BC$ cut can be rewritten as,}
\begin{multline}
	$$
	\;\;\;\;\;\;\;\;\;\;\;\;\;\;\;\;\;\;\;\;\;\;\;\;\;\;\;\;\;\;\;\;\;\;\;\;\;\;\;\;\;\;\;\;\;\;\;\;\;\;\;\;\;\;\;\left|\phi_{1,2,3,4}\right\rangle=|0\pm2\rangle|2\pm 5\rangle,\;\left|\phi_{5,6}\right\rangle=|1\rangle|0\pm2\rangle\\
	\left|\phi_{7,8}\right\rangle=|1\rangle|1\pm 7\rangle,\;\left|\phi_{9,10,11,12}\right\rangle=|0\pm2\rangle|0\pm1\rangle\\
	\left|\phi_{13,14}\right\rangle=|1\rangle|4\pm5\rangle,\;\left|\phi_{15,16}\right\rangle=|1\pm2\rangle|8\rangle\\
	\left|\phi_{17}\right\rangle=|2\rangle|3+4+6+7\rangle,\;\left|\phi_{18}\right\rangle=|2\rangle|3+4-6-7\rangle\\
	\left|\phi_{19}\right\rangle=|2\rangle|3-4+6-7\rangle,\;\left|\phi_{20}\right\rangle=|2\rangle|3-4-6+7\rangle\\
	\left|\phi_{21,22}\right\rangle=|0\rangle|6\pm 8\rangle,\;\left|\phi_{23,24}\right\rangle=|0\rangle|4\pm7\rangle\\
	\left|\phi_{25,26}\right\rangle=|0\pm1\rangle|3\rangle\;\;\;\;\;\;\;\;\;\;\;\;\;\;\;\;\;\;\;\;\;\;\;\;\;\;\;\;\;\;\;\;\;\;\;\;\;\;\;\;\;\;\;\;\;\;\;\;\;\;\;\;\;\;\;\;\;\;\;\;\;\;\;\;\;
	$$
	\label{15}
\end{multline}
By choosing appropriate states from (\ref{15}) we can find out the $9\times9$ matrix elements corresponding to the measurement of $BC$,
$$
\begin{array}{|c|c||c|c|}
	\hline \text { states } & \text { elements } & \text { states } & \text { elements } \\
	\hline \hline|0+2\rangle|2+5\rangle,|0+2\rangle|2-5\rangle & a_{22}=a_{55} & |1\rangle|0+2\rangle,|1\rangle|0-2\rangle & a_{00}=a_{22} \\
	\hline|1\rangle|1+7\rangle,|1\rangle|1-7\rangle & a_{11}=a_{77} & |0+2\rangle|0+1\rangle,|0+2\rangle|0-1\rangle & a_{00}=a_{11}\\
	\hline|1\rangle|4+5\rangle,|1\rangle|4-5\rangle & a_{44}=a_{55} & |0\rangle|6+8\rangle,|0\rangle|6-8\rangle & a_{66}=a_{88}\\
	\hline|0\rangle|4+7\rangle,|0\rangle|4-7\rangle & a_{44}=a_{77} &  & \\
	\hline
\end{array}
$$
$$
\begin{array}{|c|c|c|}
	\hline \text { Serial Number } & \text { States } & \text { Elements }\\
	\hline \hline(1) & |0+1\rangle|3\rangle,|0\rangle|4\pm7\rangle & a_{34}=a_{43}=a_{37}=a_{73}=0 \\
	\hline(2) & |0+1\rangle|3\rangle,|1+2\rangle|8\rangle & a_{38}=a_{83}=0\\
	\hline(3) & |0+1\rangle|3\rangle,|0\rangle|6+8\rangle & a_{36}=a_{63}=0\\
	\hline(4) & |0+1\rangle|3\rangle,|1\rangle|1+7\rangle & a_{31}=a_{13}=0\\
	\hline(5) & |0+1\rangle|3\rangle,|0+2\rangle|0+1\rangle & a_{30}=a_{03}=0 \\
	\hline(6) & |0+1\rangle|3\rangle,|1\rangle|0+2\rangle & a_{32}=a_{23}=0\\
	\hline(7) & |0+1\rangle|3\rangle,|1\rangle|4+5\rangle & a_{35}=a_{53}=0 \\
	\hline(8) & |1\rangle|4\pm5\rangle,|1+2\rangle|8\rangle & a_{84}=a_{48}=a_{85}=a_{58}=0\\
	\hline(9) & |0+2\rangle|2+5\rangle,|1+2\rangle|8\rangle & a_{82}=a_{28}=0\\
	\hline(10) & |0+2\rangle|0\pm1\rangle,|1+2\rangle|8\rangle & a_{80}=a_{08}=a_{81}=a_{18}=0\\
	\hline(11) & |1\rangle|1+7\rangle,|1+2\rangle|8\rangle & a_{87}=a_{78}=0\\
	\hline(12) & |2\rangle|3+4+6+7\rangle,|1+2\rangle|8\rangle & a_{86}=a_{68}=0\\
	\hline(13) & |0\rangle|4\pm7\rangle,|0\rangle|6+8\rangle & a_{76}=a_{67}=a_{64}=a_{46}=0\\
	\hline(14) & |0+2\rangle|0\pm1\rangle,|0\rangle|4\pm7\rangle & a_{70}=a_{07}=a_{71}=a_{17}=a_{40}=a_{04}=a_{41}=a_{14}=0\\
	\hline(15) & |0+2\rangle|2\pm5\rangle,|0\rangle|4\pm7\rangle & a_{72}=a_{27}=a_{75}=a_{57}=a_{42}=a_{24}=a_{45}=a_{54}=0\\
	\hline(16) & |0+2\rangle|0\pm1\rangle,|0\rangle|6+8\rangle & a_{60}=a_{06}=a_{61}=a_{16}=0\\
	\hline(17) & |0+2\rangle|2\pm5\rangle,|0\rangle|6+8\rangle & a_{62}=a_{26}=a_{65}=a_{56}=0\\
	\hline(18) & |1\rangle|0\pm2\rangle,|1\rangle|4+5\rangle & a_{50}=a_{05}=a_{52}=a_{25}=0\\
	\hline(19) & |0+2\rangle|2\pm5\rangle,|0+2\rangle|0\pm1\rangle & a_{51}=a_{15}=a_{20}=a_{02}=a_{21}=a_{12}=0\\
	\hline(20) & |1\rangle|0+2\rangle,|1\rangle|1+7\rangle & a_{10}=a_{01}=0\\
	\hline
\end{array}
$$
the remaining elements are $$a_{74},a_{47}.
$$
Now if we take the states $|2\rangle|3-4-6+7\rangle,|2\rangle|3+4-6-7\rangle$ and since The measurement must leave the postmeasurement states mutually orthogonal. Therefore we have,\\
\[\begin{aligned}&\left\langle\phi_{i}\left| \mathbb{I} \otimes E_{m}\right| \phi_{j}\right\rangle=0
	\\&\quad \implies \langle 2||2\rangle\langle|3-4-6+7|E_{m}|3+4-6-7\rangle=0 \\&\quad \implies a_{33}+a_{34}-a_{36}-a_{37}-a_{43}-a_{44}+a_{46}+a_{47}-a_{63}-a_{64}+a_{66}+a_{67}+a_{73}+a_{74}-a_{76}-a_{77}=0 \\&\quad \implies
	a_{47}+a_{74}=0
	\\&\quad \implies
	2a_{74}=0
	\\&\quad \implies
	a_{74}=0
\end{aligned}\]
[All the remaining terms of the equation becomes canceled out by using the conditions $a_{ij}=a_{ji}=0, \;\;\;i,j\neq4,7$ \;\;\;and $a_{ii}=a_{jj} \;\;\;\;\forall i,j.$]\\
Therefore we have $a_{74}=a_{47}=0$.\\
Thus,  $BC$ cannot start a nontrivial nondisturbing measurement. Also we show previously that the measurement corresponding to $A$ must be trivial. Hence the above set of states is locally irreducible in $A|BC$ bipartition.\\\\
{\bfseries The set of states in $AC|B$ cut can be rewritten as,}
\begin{multline}
	$$
\;\;\;\;\;\;\;\;\;\;\;\;\;\;\;\;\;\;\;\;\;\;\;\;\;\;\;\;\;\;\;\;\;\;\;\;\;\;\;\;\;\;\;\;\;\;\;\;\;\;\;\;\;\;\left|\phi_{1,2,3,4}\right\rangle=|2\pm8\rangle|0\pm1\rangle,\;\left|\phi_{5,6}\right\rangle=|3\pm5\rangle|0\rangle\\
	\left|\phi_{7,8}\right\rangle=|4\rangle|0\pm 2\rangle,\;\left|\phi_{9}\right\rangle=|0+1+6+7\rangle|0\rangle\\
	\left|\phi_{10}\right\rangle=|0+1-6-7\rangle|0\rangle,\;\left|\phi_{11}\right\rangle=|0-1+6-7\pm2\rangle|0\rangle\\
	\left|\phi_{12}\right\rangle=|0-1-6+7\rangle|0\rangle,\;\left|\phi_{13,14}\right\rangle=|4\pm5\rangle|1\rangle\\
	\left|\phi_{15,16}\right\rangle=|5\pm8\rangle|2\rangle,\;\left|\phi_{17,18,19,20}\right\rangle=|6\pm7\rangle|1\pm2\rangle\\
	\left|\phi_{21,22}\right\rangle=|0\pm2\rangle|2\rangle,\;\left|\phi_{23,24}\right\rangle=|1\rangle|1\pm2\rangle\\
	\left|\phi_{25,26}\right\rangle=|0\pm3\rangle|1\rangle\;\;\;\;\;\;\;\;\;\;\;\;\;\;\;\;\;\;\;\;\;\;\;\;\;\;\;\;\;\;\;\;\;\;\;\;\;\;\;\;\;\;\;\;\;\;\;\;\;\;\;\;\;\;\;\;\;\;\;\;\;\;\;\;\;
	$$
	\label{16}
\end{multline}
By choosing appropriate states from (\ref{16}) we can find out the $9\times9$ matrix elements corresponding to the measurement of $AC$,
$$
\begin{array}{|c|c||c|c|}
	\hline \text { states } & \text { elements } & \text { states } & \text { elements } \\
	\hline \hline|2+8\rangle|0+1\rangle,|2-8\rangle|0+1\rangle & a_{22}=a_{88} & |3+5\rangle|0\rangle,|3-5\rangle|0\rangle & a_{33}=a_{55} \\
	\hline|4+5\rangle|1\rangle,|4-5\rangle|1\rangle & a_{44}=a_{55} & |5+8\rangle|2\rangle,|5-8\rangle|2\rangle & a_{55}=a_{88}\\
	\hline|6+7\rangle|1+2\rangle,|6-7\rangle|1+2\rangle & a_{66}=a_{77} & |0+2\rangle|2\rangle,|0-2\rangle|2\rangle & a_{00}=a_{22}\\	\hline|0+3\rangle|1\rangle,|0-3\rangle|1\rangle & a_{00}=a_{33} &  & \\
	\hline
\end{array}
$$
$$
\begin{array}{|c|c|c|}
	\hline \text { Serial Number } & \text { States } & \text { Elements }\\
	\hline \hline(1) & |0+2\rangle|2\rangle,|1\rangle|1+2\rangle & a_{10}=a_{01}=a_{12}=a_{21}=0 \\
	\hline(2) & |1\rangle|1+2\rangle,|0+3\rangle|1\rangle & a_{13}=a_{31}=0\\
	\hline(3) & |4\pm5\rangle|1\rangle,|1\rangle|1+2\rangle & a_{14}=a_{41}=a_{15}=a_{51}=0\\
	\hline(4) & |6\pm7\rangle|1+2\rangle,|1\rangle|1+2\rangle & a_{16}=a_{61}=a_{17}=a_{71}=0\\
	\hline(5) & |5+8\rangle|2\rangle,|1\rangle|1+2\rangle & a_{18}=a_{81}=0 \\
	\hline(6) & |4\rangle|0+2\rangle,|0\pm2\rangle|2\rangle & a_{40}=a_{04}=a_{42}=a_{24}=0\\
	\hline(7) & |3\pm5\rangle|0\rangle,|4\rangle|0+2\rangle & a_{43}=a_{34}=a_{45}=a_{54}=0 \\
	\hline(8) & |4\rangle|0+2\rangle,|6\pm7\rangle|1+2\rangle & a_{46}=a_{64}=a_{47}=a_{74}=0\\
	\hline(9) & |4\rangle|0+2\rangle,|5+8\rangle|2\rangle & a_{48}=a_{84}=0\\
	\hline(10) & |2\pm8\rangle|0+1\rangle,|3\pm5\rangle|0\rangle & a_{23}=a_{32}=a_{25}=a_{52}=a_{83}=a_{38}=a_{85}=a_{58}=0\\
	\hline(11) & |2\pm8\rangle|0+1\rangle,|6\pm7\rangle|1+2\rangle & a_{26}=a_{62}=a_{27}=a_{72}=a_{86}=a_{68}=a_{87}=a_{78}=0\\
	\hline(12) & |5\pm8\rangle|2\rangle,|0\pm2\rangle|2\rangle & a_{50}=a_{05}=a_{80}=a_{08}=a_{28}=a_{82}=0\\
	\hline(13) & |2+8\rangle|0+1\rangle,|0+3\rangle|1\rangle & a_{20}=a_{02}=0\\
	\hline(14) & |5+8\rangle|2\rangle,|6\pm7\rangle|1+2\rangle & a_{56}=a_{65}=a_{57}=a_{75}=0\\
	\hline(15) & |4+5\rangle|1\rangle,|0+3\rangle|1\rangle & a_{53}=a_{35}=0\\
	\hline(16) & |6\pm7\rangle|1+2\rangle,|0\pm3\rangle|1\rangle & a_{60}=a_{06}=a_{63}=a_{36}=a_{70}=a_{07}=a_{73}=a_{37}=0\\
	\hline(17) & |3+5\rangle|0\rangle,|0+1+6+7\rangle|0\rangle & a_{30}=a_{03}=0\\
	\hline
\end{array}
$$
the remaining elements are $$a_{67},a_{76}.
$$
Now if we take the states $|2\rangle|0+1+6+7\rangle,|2\rangle|0+1-6-7\rangle$ and since The measurement must leave the postmeasurement states mutually orthogonal. Therefore we have,\\
\[\begin{aligned}&\left\langle\phi_{i}\left|E_{m} \otimes \mathbb{I}\right| \phi_{j}\right\rangle=0
	\\&\quad \implies\langle|0+1+6+7|E_{m}|0+1-6-7\rangle \langle 0||0\rangle=0 \\&\quad \implies a_{00}+a_{01}-a_{06}-a_{07}+a_{10}+a_{11}-a_{16}-a_{17}+a_{60}+a_{61}-a_{66}-a_{67}+a_{70}+a_{71}-a_{76}-a_{77}=0 \\&\quad \implies
	-a_{67}-a_{76}=0
	\\&\quad \implies
	-2a_{76}=0
	\\&\quad \implies
	a_{76}=0
\end{aligned}\]
[All the remaining terms of the equation becomes canceled out by using the conditions $a_{ij}=a_{ji}=0, \;\;\;i,j\neq 6,7$ \;\;\;and $a_{ii}=a_{jj} \;\;\;\;\forall i,j.$]\\
Therefore we have $a_{76}=a_{67}=0$.\\
Thus,  $AC$ cannot start a nontrivial nondisturbing measurement. Also we show previously that the measurement corresponding to $B$ must be trivial. Hence the above set of states is locally irreducible in $AC|B$ bipartition.
 Hence the set of states (\ref{A}) is irreducible in all the bipartition also. This completes the proof. \(\blacksquare \)\\
\end{widetext}
\section*{Appendix B: proof of Proposition 8.}
 Firstly let us assume that Bob will teleport his system to Alice by using $\log_2 3$ ebit of entanglement as a resource. Therefore the set of states in $AB|C$ cut can be rewritten as,
 \begin{multline}
 	$$
 	\;\;\;\;\;\;\;\;\;\;\left|\phi_{1,2}\right\rangle=|1\rangle|0\pm1\rangle,\;\;\;\left|\phi_{3,4}\right\rangle=|0\pm3\rangle|1\rangle\\
 	\left|\phi_{5,6}\right\rangle=|3\pm 4\rangle|0\rangle,\;\;\;\;\left|\phi_{7,8}\right\rangle=|2\rangle|0\pm2\rangle\;\;\\
 	\left|\phi_{9,10}\right\rangle=|0\pm6\rangle|2\rangle,\;\;\;\;\;\left|\phi_{11,12}\right\rangle=|6\pm8\rangle|0\rangle\\
 	\left|\phi_{13,14}\right\rangle=|5\rangle|0 \pm1\rangle,\;\;\;\;\left|\phi_{15,16}\right\rangle=|1\pm4\rangle|2\rangle\\
 	\left|\phi_{17,18}\right\rangle=|6\pm7\rangle|1\rangle,\; \quad\left|\phi_{19,20}\right\rangle=|7\rangle|0\pm2\rangle\\
 	\;\;\;\;\;\;\;\;\left|\phi_{21,22}\right\rangle=|2\pm8\rangle|1\rangle,\;\quad\left|\phi_{23,24}\right\rangle=|3\pm5\rangle|2\rangle
 	$$
 	\label{D}
 \end{multline}
Now let us assume that a two qubit Bell state shared between two parties $AB$ and $C$ be $|\psi\rangle_{ab|c}$. Therefore the initial states shared among them is $$\left|\phi\right\rangle_{AB|C}\otimes\left|\phi\right\rangle_{ab|c}$$
Where $\left|\phi\right\rangle$ is one of the state from (\ref{D}).\\
Step $1 .$ Charlie performs a measurement
$$
\begin{aligned}
	\mathcal{C} & \equiv\left\{M:=\mathbb{P}\left[(|0\rangle,|1\rangle)_{C} ;|0\rangle_{c}\right]+\mathbb{P}\left[|2\rangle_{C} ;|1\rangle_{c}\right]\right.\\
	\bar{M} &:=\mathbb{I}-M\}
\end{aligned}
$$
Suppose the outcomes corresponding to $M$ click. The resulting postmeasurement states is therefore\\
\begin{multline*}
	$$
	\;\left|\phi_{1,2}\right\rangle \rightarrow\left|1\rangle_{AB}|0 \pm 1\rangle_{C}|00\rangle_{ab|c}\right.,\\
	\left|\phi_{3,4}\right\rangle \rightarrow\left|0\pm3\rangle_{AB}|1\rangle_{C}|00\rangle_{ab|c}\right.,\;\;\;\;\;\;\;\;\;\;\;\;\;\;\;\;\;\;\;\;\;\;\;\;\;\;\;\;\;\;\;\\
	\left|\phi_{5,6}\right\rangle \rightarrow\left|3\pm4\rangle_{AB}|0\rangle_{C}|00\rangle_{ab|c}\right.,\;\;\;\;\;\;\;\;\;\;\;\;\;\;\;\;\;\;\;\;\;\;\;\;\;\;\;\;\;\;\;\\
	\left|\phi_{7,8}\right\rangle \rightarrow\left|2\rangle_{AB}|0\rangle_{C}|00\rangle_{ab|c}+|2\rangle_{AB}|2\rangle_{C}|11\rangle_{ab|c}\right.,\;\;\;\;\;\;\\
	\left|\phi_{9,10}\right\rangle \rightarrow\left|0\pm6\rangle_{AB}|2\rangle_{C}|11\rangle_{ab|c}\right.,\;\;\;\;\;\;\;\;\;\;\;\;\;\;\;\;\;\;\;\;\;\;\;\;\;\;\;\;\;\\
	\left|\phi_{11,12}\right\rangle \rightarrow\left|6\pm8\rangle_{AB}|0\rangle_{C}|00\rangle_{ab|c}\right.,\;\;\;\;\;\;\;\;\;\;\;\;\;\;\;\;\;\;\;\;\;\;\;\;\;\;\;\;\\
	\left|\phi_{13,14}\right\rangle \rightarrow\left|5\rangle_{AB}|0\pm1\rangle_{C}|00\rangle_{ab|c}\right.,\;\;\;\;\;\;\;\;\;\;\;\;\;\;\;\;\;\;\;\;\;\;\;\;\;\;\;\;\\
	\left|\phi_{15,16}\right\rangle \rightarrow\left|1\pm4\rangle_{AB}|2\rangle_{C}|11\rangle_{ab|c}\right.,\;\;\;\;\;\;\;\;\;\;\;\;\;\;\;\;\;\;\;\;\;\;\;\;\;\;\;\;\\
	\left|\phi_{17,18}\right\rangle \rightarrow\left|6\pm7\rangle_{AB}|1\rangle_{C}|00\rangle_{ab|c}\right.,\;\;\;\;\;\;\;\;\;\;\;\;\;\;\;\;\;\;\;\;\;\;\;\;\;\;\;\;\\
	\left|\phi_{19,20}\right\rangle \rightarrow\left|7\rangle_{AB}|0\rangle_{C}|00\rangle_{ab|c}+|7\rangle_{AB}|2\rangle_{C}|11\rangle_{ab|c}\right.\;\;\;\;\\
	\left|\phi_{21,22}\right\rangle \rightarrow\left|2\pm8\rangle_{AB}|1\rangle_{C}|00\rangle_{ab|c}\right.,\;\;\;\;\;\;\;\;\;\;\;\;\;\;\;\;\;\;\;\;\;\;\;\;\;\;\;\\
	\left|\phi_{23,24}\right\rangle \rightarrow\left|3\pm5\rangle_{AB}|2\rangle_{C}|11\rangle_{ab|c}\right..\;\;\;\;\;\;\;\;\;\;\;\;\;\;\;\;\;\;\;\;\;\;\;\;\;\;\;\\
	$$
\end{multline*}
Step $2 .$ $AB$ performs ten outcome projective measurement,\\\\
$N_{1}:=\mathbb{P}[(|0\rangle_{AB},|3\rangle_{AB},|4\rangle_{AB}) ;|0\rangle_{ab}]$\\
$N_{2}:=\mathbb{P}[|3+5\rangle_{AB} ;|1\rangle_{ab}]
$, $N_{3}:=\mathbb{P}[|3-5\rangle_{AB} ;|1\rangle_{ab}]
$\\
$N_{4}:=\mathbb{P}[|1+4\rangle_{AB} ;|1\rangle_{ab}]
$, $N_{5}:=\mathbb{P}[|1-4\rangle_{AB} ;|1\rangle_{ab}]
$\\
$N_{6}:=\mathbb{P}[|0+6\rangle_{AB} ;|1\rangle_{ab}]
$, $N_{7}:=\mathbb{P}[|0-6\rangle_{AB} ;|1\rangle_{ab}]
$\\
$N_{8}:=\mathbb{P}[|1\rangle_{AB} ;|0\rangle_{ab}]
$,  $N_{9}:=\mathbb{P}[|5\rangle_{AB} ;|0\rangle_{ab}]
$\\
$N_{10}=\mathbb{I}-{\sum_{i=1}^{9} N_{i}}$\\\\
If the outcomes corresponding to $N_1$ click. The resulting post measurement states are therefore $\left|0\pm3\rangle_{AB}|1\rangle_{C}|00\rangle_{ab|c}\right.$, $\left|3\pm4\rangle_{AB}|0\rangle_{C}|00\rangle_{ab|c}\right.$ which can be easily distinguished by $AB$ easily after projecting onto $|0\rangle_{C}$, $|1\rangle_{C}$ by Charlie. If the outcomes corresponding to $N_2$ and $N_3$ click, then they discriminate $\left|\phi_{23}\right\rangle$, $\left|\phi_{24}\right\rangle$ respectively. Also if the outcomes corresponding to $N_4$ and $N_5$ click, then they discriminate $\left|\phi_{15}\right\rangle$, $\left|\phi_{16}\right\rangle$ respectively. Similarly they discriminate $\left|\phi_{9}\right\rangle$, $\left|\phi_{10}\right\rangle$ by projecting onto $N_6$ and $N_7$ respectively. Now if the outcome $N_8$ click, the post measurement states becomes $\left|\phi_{1,2}\right\rangle$ which can be easily distinguished by Charlie by projecting onto $|0\pm1\rangle_C$. Similarly if the outcome $N_9$ click, the post measurement states becomes $\left|\phi_{13,14}\right\rangle$ which can be easily distinguished by Charlie by projecting onto $|0\pm1\rangle_C$. \\\\
The remaining states are isolated corresponding to the outcome $N_{10}$. Which are
$\left|2\rangle_{AB}|0\rangle_{C}|00\rangle_{ab|c}+|2\rangle_{AB}|2\rangle_{C}|11\rangle_{ab|c}\right.$,\\ $\left|7\rangle_{AB}|0\rangle_{C}|00\rangle_{ab|c}+|7\rangle_{AB}|2\rangle_{C}|11\rangle_{ab|c}\right.$,\\ $\left|6\pm8\rangle_{AB}|0\rangle_{C}|00\rangle_{ab|c}\right.$, $\left|6\pm7\rangle_{AB}|1\rangle_{C}|00\rangle_{ab|c}\right.$,\\ $\left|2\pm8\rangle_{AB}|1\rangle_{C}|00\rangle_{ab|c}\right.$.\\\\
Step $3 .$ Charlie performs a measurement
$$
\begin{aligned}
	\mathcal{C^\prime} & \equiv\left\{Q:=\mathbb{P}\left[|0\rangle_{C} ;|0\rangle_{c}\right]+\mathbb{P}\left[|2\rangle_{C} ;|1\rangle_{c}\right]\right.\\
	\bar{Q} &:=\mathbb{I}-Q\}
\end{aligned}
$$
If the outcomes corresponding to $Q$ click. The resulting post measurement states are therefore $\left|2\rangle_{AB}|0\rangle_{C}|00\rangle_{ab|c}+|2\rangle_{AB}|2\rangle_{C}|11\rangle_{ab|c}\right.$, $\left|7\rangle_{AB}|0\rangle_{C}|00\rangle_{ab|c}+|7\rangle_{AB}|2\rangle_{C}|11\rangle_{ab|c}\right.$, $\left|6\pm8\rangle_{AB}|0\rangle_{C}|00\rangle_{ab|c}\right.$. Afterthat $AB$ together makes four outcome projective measurement $R_1=|2\rangle_{AB}\langle2|\otimes(|0\rangle_{ab}\langle0|+|1\rangle_{ab}\langle1|)$, $R_2=|7\rangle_{AB}\langle7|\otimes(|0\rangle_{ab}\langle0|+|1\rangle_{ab}\langle1|)$, $R_3=|6+8\rangle_{AB}\langle6+8|\otimes|0\rangle_{ab}\langle0|$, $R_4=|6-8\rangle_{AB}\langle6+8|\otimes|0\rangle_{ab}\langle0|$. The outcome $R_1$ corresponds two states $\left|\phi_{7,8}\right\rangle$ which can be easily distinguished by Walgate.et.al \cite{Walgate2000} result. Similarly $\left|\phi_{19,20}\right\rangle$ can be distinguished if $R_2$ occur. The outcomes $R_3$ and $R_4$ omits $\left|\phi_{11}\right\rangle$ and $\left|\phi_{12}\right\rangle$ respectively.\\
Next if the outcome $\bar{Q}$ occur, it isolates four states $\left|6\pm7\rangle_{AB}|1\rangle_{C}|00\rangle_{ab|c}\right.$, $\left|2\pm8\rangle_{AB}|1\rangle_{C}|00\rangle_{ab|c}\right.$, which can be distinguished by $AB$ by projecting onto $|6\pm7\rangle_{AB}$ and $|2\pm8\rangle_{AB}$, respectively. Hence this completes the proof. \(\blacksquare \)

\section*{Appendix C:  (1+$\log_23$) ebit of entanglement is not sufficient to distinguish the set (\ref{A}).}
Firstly let us assume that Bob will teleport his system to Alice by using $\log_23$ ebit of entanglement as a resource. Therefore the set of states in $AB|C$ cut can be rewritten as,
\begin{multline}
	$$
	\left|\phi_{1}\right\rangle=|0+6+1+7\rangle|2\rangle,\;\left|\phi_{2}\right\rangle=|0+6-1-7\rangle|2\rangle\\
	\left|\phi_{3}\right\rangle=|0-6+1-7\rangle|2\rangle,\;\left|\phi_{4}\right\rangle=|0-6-1+7\rangle|2\rangle\\
	\left|\phi_{5,6}\right\rangle=|3\rangle|0 \pm 2\rangle,\;\left|\phi_{7,8}\right\rangle=|3 \pm 5\rangle|1\rangle\\
	\left|\phi_{9,10,11,12}\right\rangle=|0\pm6\rangle|0 \pm 1\rangle,\;\left|\phi_{13,14}\right\rangle=|4\rangle|1\pm2\rangle\\
	\left|\phi_{15,16}\right\rangle=|5\pm8\rangle|2\rangle,\;\left|\phi_{17,18,19,20}\right\rangle=|7 \pm 8\rangle|0\pm1\rangle\\
	\left|\phi_{21,22}\right\rangle=|2\rangle|0 \pm 2\rangle,\;\left|\phi_{23,24}\right\rangle=|1\pm 2\rangle|1\rangle\\
	\left|\phi_{25,26}\right\rangle=|1\pm4\rangle|0\rangle\;\;\;\;\;\;\;\;\;\;\;\;\;\;\;\;\;\;\;\;\;
	$$
	\label{X}
\end{multline}
Now let us assume that a two qubit Bell state shared between two parties $AB$ and $C$ be $|\psi\rangle_{ab|c}$. Therefore the initial states shared among them is $$\left|\phi\right\rangle_{AB|C}\otimes\left|\phi\right\rangle_{ab|c}$$
Where $\left|\phi\right\rangle$ is one of the state from (\ref{X}).\\
Now the possible choice of measurement of Charlie must belongs to the set of $^3P_2$ members, where $^nP_k= \frac{n!}{(n-k)!}$\\
$$
\begin{Bmatrix}
\begin{aligned}
	\mathcal{C} & \equiv\left\{M:=\mathbb{P}\left[(|0\rangle,|1\rangle)_{C} ;|0\rangle_{c}\right]+\mathbb{P}\left[|2\rangle_{C} ;|1\rangle_{c}\right]\right.\\
	\bar{M} &:=\mathbb{I}-M\}
\end{aligned}\\\\
\begin{aligned}
	\mathcal{C} & \equiv\left\{M:=\mathbb{P}\left[(|0\rangle,|2\rangle)_{C} ;|0\rangle_{c}\right]+\mathbb{P}\left[|1\rangle_{C} ;|1\rangle_{c}\right]\right.\\
	\bar{M} &:=\mathbb{I}-M\}
\end{aligned}\\\\
\begin{aligned}
	\mathcal{C} & \equiv\left\{M:=\mathbb{P}\left[(|1\rangle,|2\rangle)_{C} ;|0\rangle_{c}\right]+\mathbb{P}\left[|0\rangle_{C} ;|1\rangle_{c}\right]\right.\\
	\bar{M} &:=\mathbb{I}-M\}
\end{aligned}\\\\
\begin{aligned}
	\mathcal{C} & \equiv\left\{M:=\mathbb{P}\left[(|0\rangle,|1\rangle)_{C} ;|1\rangle_{c}\right]+\mathbb{P}\left[|2\rangle_{C} ;|0\rangle_{c}\right]\right.\\
	\bar{M} &:=\mathbb{I}-M\}
\end{aligned}\\\\
\begin{aligned}
	\mathcal{C} & \equiv\left\{M:=\mathbb{P}\left[(|0\rangle,|2\rangle)_{C} ;|1\rangle_{c}\right]+\mathbb{P}\left[|1\rangle_{C} ;|0\rangle_{c}\right]\right.\\
	\bar{M} &:=\mathbb{I}-M\}
\end{aligned}\\\\
\begin{aligned}
	\mathcal{C} & \equiv\left\{M:=\mathbb{P}\left[(|1\rangle,|2\rangle)_{C} ;|1\rangle_{c}\right]+\mathbb{P}\left[|0\rangle_{C} ;|0\rangle_{c}\right]\right.\\
	\bar{M} &:=\mathbb{I}-M\}
\end{aligned}
\end{Bmatrix}
$$
The first three choice of measurement are identical with the last three. Because they just exchanging their bits of the ancillary subsystems.  

$1.$ If Charlie performs a measurement
$$
\begin{aligned}
	\mathcal{C} & \equiv\left\{M:=\mathbb{P}\left[(|0\rangle,|1\rangle)_{C} ;|0\rangle_{c}\right]+\mathbb{P}\left[|2\rangle_{C} ;|1\rangle_{c}\right]\right.\\
	\bar{M} &:=\mathbb{I}-M\}
\end{aligned}
$$
Suppose the outcomes corresponding to $M$ click. The resulting postmeasurement states is therefore\\
\begin{multline}
	$$
	\;\;\left|\phi_{1}\right\rangle=|01+61+11+71\rangle|21\rangle,\;\left|\phi_{2}\right\rangle=|01+61-11-71\rangle|21\rangle\\
	\left|\phi_{3}\right\rangle=|01-61+11-71\rangle|21\rangle,\;\left|\phi_{4}\right\rangle=|01-61-11+71\rangle|21\rangle\\
	\left|\phi_{5,6}\right\rangle=|30\rangle|00\rangle \pm |31\rangle|21\rangle,\;\left|\phi_{7,8}\right\rangle=|30\pm 50\rangle|10\rangle\\
	\left|\phi_{9,10,11,12}\right\rangle=|00\pm60\rangle|00 \pm 10\rangle,\;\left|\phi_{13,14}\right\rangle=|40\rangle|10\rangle\pm|41\rangle|21\rangle\\
	\left|\phi_{15,16}\right\rangle=|51\pm81\rangle|21\rangle,\;\left|\phi_{17,18,19,20}\right\rangle=|70\pm80\rangle|00\pm10\rangle\\
	\left|\phi_{21,22}\right\rangle=|20\rangle|00\rangle\pm|21\rangle|0 \pm21\rangle,\;\left|\phi_{23,24}\right\rangle=|10\pm 20\rangle|10\rangle\\
	\left|\phi_{25,26}\right\rangle=|10\pm40\rangle|00\rangle\;\;\;\;\;\;\;\;\;\;\;\;\;\;\;\;\;
	$$
\end{multline}
Next $AB$ can perform maximum six outcome projective measurement,\\\\
$N_{1}:=\mathbb{P}[(|0\rangle_{AB},|6\rangle_{AB},|1\rangle_{AB},|7\rangle_{AB}) ;|1\rangle_{ab}]$\\
$N_{2}:=\mathbb{P}[|7\rangle_{AB},|8\rangle_{AB} ;|0\rangle_{ab}]
$, $N_{3}:=\mathbb{P}[|5\rangle_{AB},|8\rangle_{AB} ;|1\rangle_{ab}]
$\\
$N_{4}:=\mathbb{P}[|0\rangle_{AB},|6\rangle_{AB} ;|0\rangle_{ab}]
$,\\ $N_{5}:=\mathbb{P}[|3\rangle_{AB},|5\rangle_{AB} ;|0\rangle_{ab}]+\mathbb{P}[|3\rangle_{AB} ;|1\rangle_{ab}]
$\\
$N_{6}:=\mathbb{P}[|1\rangle_{AB},|2\rangle_{AB},|4\rangle_{AB} ;|0\rangle_{ab}]+\mathbb{P}[|4\rangle_{AB},|2\rangle_{AB} ;|1\rangle_{ab}]
$\\
But it can be checked that if the outcome $N_{6}$ occur, the set cannot be distinguished perfectly.\\
$2.$ If Charlie performs a measurement
$$
\begin{aligned}
	\mathcal{C} & \equiv\left\{M:=\mathbb{P}\left[(|0\rangle,|2\rangle)_{C} ;|0\rangle_{c}\right]+\mathbb{P}\left[|1\rangle_{C} ;|1\rangle_{c}\right]\right.\\
	\bar{M} &:=\mathbb{I}-M\}
\end{aligned}
$$
Suppose the outcomes corresponding to $M$ click. The resulting postmeasurement states is therefore\\
\begin{multline}
	$$
	\;\;\left|\phi_{1}\right\rangle=|00+60+10+70\rangle|20\rangle,\;\left|\phi_{2}\right\rangle=|00+60-10-70\rangle|20\rangle\\
	\left|\phi_{3}\right\rangle=|00-60+10-70\rangle|20\rangle,\;\left|\phi_{4}\right\rangle=|00-60-10+70\rangle|20\rangle\\
	\left|\phi_{5,6}\right\rangle=|30\rangle|00\pm20\rangle,\;\left|\phi_{7,8}\right\rangle=|31\pm 51\rangle|11\rangle\\
	\left|\phi_{9,10,11,12}\right\rangle=|00\pm60\rangle|00\rangle\pm |01\pm61\rangle|11\rangle,\\\left|\phi_{13,14}\right\rangle=|41\rangle|11\rangle\pm|40\rangle|20\rangle,
	\left|\phi_{15,16}\right\rangle=|50\pm80\rangle|20\rangle,\;\\\left|\phi_{17,18,19,20}\right\rangle=|70\pm80\rangle|00\rangle\pm|71\pm81\rangle|11\rangle\\
	\left|\phi_{21,22}\right\rangle=|20\rangle|00\pm20\rangle,\;\left|\phi_{23,24}\right\rangle=|11\pm 21\rangle|11\rangle\\
	\left|\phi_{25,26}\right\rangle=|10\pm40\rangle|00\rangle\;\;\;\;\;\;\;\;\;\;\;\;\;\;\;\;\;
	$$
\end{multline}
Next $AB$ can perform maximum four outcome projective measurement,\\\\
$N_{1}:=\mathbb{P}[|1\rangle_{AB},|2\rangle_{AB} ;|1\rangle_{ab}]
$\\
$N_{2}:=\mathbb{P}[|2\rangle_{AB}, ;|0\rangle_{ab}]
$,\\ $N_{3}:=\mathbb{P}[|3\rangle_{AB},|5\rangle_{AB} ;|1\rangle_{ab}]+\mathbb{P}[|3\rangle_{AB} ;|0\rangle_{ab}]
$\\
$N_{4}:=\mathbb{P}[|0\rangle_{AB},|1\rangle_{AB},|4\rangle_{AB},|5\rangle_{AB},|6\rangle_{AB},|7\rangle_{AB},|8\rangle_{AB} ;|0\rangle_{ab}]$+\\$\mathbb{P}[|0\rangle_{AB},|4\rangle_{AB},|6\rangle_{AB},|7\rangle_{AB},|8\rangle_{AB} ;|1\rangle_{ab}]
$\\
But it can be checked that if the outcome $N_{4}$ occur, the set cannot be distinguished perfectly.\\
$3.$ If Charlie performs a measurement
$$
\begin{aligned}
	\mathcal{C} & \equiv\left\{M:=\mathbb{P}\left[(|1\rangle,|2\rangle)_{C} ;|0\rangle_{c}\right]+\mathbb{P}\left[|0\rangle_{C} ;|1\rangle_{c}\right]\right.\\
	\bar{M} &:=\mathbb{I}-M\}
\end{aligned}
$$
Suppose the outcomes corresponding to $M$ click. The resulting postmeasurement states is therefore\\
\begin{multline}
	$$
	\;\;\left|\phi_{1}\right\rangle=|00+60+10+70\rangle|20\rangle,\;\left|\phi_{2}\right\rangle=|00+60-10-70\rangle|20\rangle\\
	\left|\phi_{3}\right\rangle=|00-60+10-70\rangle|20\rangle,\;\left|\phi_{4}\right\rangle=|00-60-10+70\rangle|20\rangle\\
	\left|\phi_{5,6}\right\rangle=|31\rangle|01\rangle\pm|30\rangle|20\rangle,\;\left|\phi_{7,8}\right\rangle=|30\pm 50\rangle|10\rangle\\
	\left|\phi_{9,10,11,12}\right\rangle=|01\pm61\rangle|01\rangle\pm |00\pm60\rangle|10\rangle,\\\left|\phi_{13,14}\right\rangle=|40\rangle|10\pm20\rangle,
	\left|\phi_{15,16}\right\rangle=|50\pm80\rangle|20\rangle,\;\\\left|\phi_{17,18,19,20}\right\rangle=|71\pm81\rangle|01\rangle\pm|70\pm80\rangle|10\rangle\\
	\left|\phi_{21,22}\right\rangle=|21\rangle|01\rangle\pm|20\rangle|20\rangle,\;\left|\phi_{23,24}\right\rangle=|10\pm 20\rangle|10\rangle\\
	\left|\phi_{25,26}\right\rangle=|11\pm41\rangle|01\rangle\;\;\;\;\;\;\;\;\;\;\;\;\;\;\;\;\;
	$$
\end{multline}
Next $AB$ can perform maximum three outcome projective measurement,\\\\
$N_{1}:=\mathbb{P}[|1\rangle_{AB},|4\rangle_{AB} ;|1\rangle_{ab}]
$\\
$N_{2}:=\mathbb{P}[|4\rangle_{AB}, ;|0\rangle_{ab}]
$,\\
$N_{4}:=\mathbb{P}[|0\rangle_{AB},|1\rangle_{AB},|2\rangle_{AB},|3\rangle_{AB},|4\rangle_{AB},|5\rangle_{AB},\\\;\;\;\;\;\;\;|6\rangle_{AB},|7\rangle_{AB},|8\rangle_{AB} ;|0\rangle_{ab}]$+\\$\mathbb{P}[|0\rangle_{AB},|2\rangle_{AB},|3\rangle_{AB},|6\rangle_{AB},|7\rangle_{AB},|8\rangle_{AB} ;|1\rangle_{ab}]
$\\
But it can be checked that if the outcome $N_{3}$ occur, the set cannot be distinguished perfectly.\\ 
Therefore the set is indistinguishable if Charlie goes first. Since the dimension of the subsystem $AB$ is large and as it contains more twisted subsystem than Charlie, it is easy to checked that the set of states is also indistinguishable if $AB$ goes first with any choice of measurement. The measurement choice of $AB$ must belongs to the set of $\frac{2^9-1}{2}=255$ members. It is not very hard to see that for any choice of measurement of $AB$ the set (\ref{X}) is indistinguishable. The similar things will happen for other two bipartition $AC|B$ and $BC|A$ also.\\
Therefore (1+$\log_2 3$) ebit of entanglement is not sufficient for discriminating the set of states (\ref{A}).

\end{document}